\documentclass{aa}  
\usepackage{natbib,float}
\usepackage{graphicx,booktabs}
\usepackage{txfonts}
\newcommand{\brg}{Br$\gamma$}
\newcommand{\hhs}{H$_2$~1$-$0~S(1)}

\newcommand{\co}{CO(3$-$2)}
\newcommand{\coo}{CO(2$-$1)}
\newcommand{\hii}{\ion{H}{ii}}
\newcommand{\msun}{$M_{\odot}$}

\newcommand{\lsun}{$L_{\odot}$}
\newcommand{\kms}{km~s$^{-1}$}
\newcommand{\pcc}{PCC~1}

\begin{document} 

   \title{The impact of a massive star cluster on its surrounding matter in the Antennae overlap region}

   \author{C.~N.~Herrera\inst{1,2}\fnmsep\thanks{herrera@iram.fr}
          \and
          F.~Boulanger\inst{3}
                    }

   \institute{Institut de Radioastronomie Millim\'etrique, 300 rue de la Piscine, Domaine Universitaire, 38406 Saint-Martin-d'H\`eres, France
             \and
             National Astronomical Observatory of Japan (NAOJ), 2-21-1 Osawa, Mitaka, Tokyo 181-8588, Japan
             \and
             Institut d'Astrophysique Spatiale, UMR 8617 CNRS, Universit\'e Paris-Sud 11,
             91405 Cedex Orsay, France 
             }

  \date{Received XX XX, 2016; accepted 31 Dec, 2016}

  \abstract
  {
 Super star clusters (SSCs), likely the progenitors of globular clusters, are one of the most extreme forms of star formation. Understanding how SSCs form is an observational challenge. Theoretical studies establish that, to form such clusters, the dynamical timescale of their parent clouds has to be shorter than the timescale of the  disruption of their parent clouds by stellar feedback. However, due to insufficient observational support, it is still unclear how feedback from SSCs acts on their surrounding matter. Studying feedback in SSCs is essential to understand how such clusters form.
Based on ALMA and VLT observations, we study this process in a SSC in the overlap region of the Antennae galaxies (22 Mpc), a spectacular example of a burst of star formation triggered by the encounter of two galaxies. 
We analyze a unique massive ($\sim$10$^7$~M$_{\odot}$) and young (1--3.5~Myr) SSC, still associated with compact molecular and ionized gas emission, which suggest that it may still be embedded in its parent molecular cloud. 
The cluster has two CO velocity components, a low velocity one spatially associated with the cluster and a high velocity one distributed in a bubble-like shape around the cluster. Our results on the low velocity component suggest that this gas did not participate in the formation of the SSC. We propose that most of the parent cloud has already been blown away, accelerated at the early stages of the SSC evolution by radiation pressure, in a timescale $\sim$1 Myr. The high velocity component may trace outflowing molecular gas from the parent cloud. Supporting evidence is found in shock heated H$_2$ gas and escaping Br$\gamma$ gas associated with this component. The low velocity component may be gas that was near the SSC when it formed but not part of its parent cloud or clumps that migrated from the SGMC environment. This gas would be dispersed by stellar winds and supernova explosions. The existing data is inconclusive as to whether the cluster is bound and will evolve as a globular cluster.  Within $\sim$100 pc from the cluster, we estimate a lower limit for the SFE of 17\%, smaller than the theoretical limit of 30\%  needed to form a bound cluster. 
 Further higher spatial resolution observations are needed to test and quantify our proposed scenario.}

   \keywords{galaxies: individual: Antennae -- galaxies: ISM}
\titlerunning{Disruption of a cloud driven by stellar feedback and turbulence} 
   \maketitle


\section{Introduction}\label{sec:intro}

 Super star clusters (SSCs), potentially the progenitors of globular clusters \citep[e.g.,][]{portegies10}, are one of the most extreme forms of star formation. They are compact (a few parsec size) and massive ($>$10$^5$~\msun) star clusters. SSCs are found in galaxy interactions \citep[e.g.,][]{whitmore95}, starburst dwarf galaxies \citep[e.g.,][]{turner98,beck02} as well as in the Milky Way \citep{clark05}.  The number of such objects greatly increases in galaxy interactions and mergers, common phenomena in the Universe. Therefore, SSCs must be ubiquitous in the Universe. 

Understanding how SSCs form and early evolve is observationally challenging, thus it is mainly based on theoretical work. Theoretical studies suggest that high external pressures ($\sim10^7$~k$_{\rm B}-10^8$~k$_{\rm B}$~cm$^{-3}$~K) are needed to form such compact and massive clusters, condition typically found in galaxy interactions \citep{elmegreen97,ashman01}. The absence of rotational shear in interacting and irregular galaxies may be an additional contribution to SSC formation, which in spirals may prevent giant molecular clouds (GMCs) to collapse into dense clusters  \citep{weidner10}. High external pressures can accelerate the collapse of molecular clouds, enhancing the star formation efficiency (SFE, the fraction of gas which turns into stars). If the local ($\lesssim$~1~pc) SFE is high($\gtrsim30-50$\%), bound cluster formation will occur  \citep{elmegreen97,ashman01}.  Recent hydrodynamical models indicate that in dense star forming regions even if the global SFE in a GMC is below 30\%, the local SFE can be higher than 50\%, allowing the formation of bound clusters \citep{kruijssen12}. However, observational evidence constraining the SFE in young massive clusters is difficult to obtain. For instance, the SFE of a young ($\sim$ 1~Myr) SSC in the dwarf galaxy NGC~5253 is estimated to be 50\% \citep{turner15} or $\sim$20\% \citep{smith16}, depending on the assumed initial mass function.

High local SFE will occur if the local dynamical timescale of the parent molecular cloud (cluster formation timescale) is shorter than the timescale of disruptive processes \citep[i.e.][]{portegies10}. The most invoked disruptive process is the {\it stellar feedback}, i.e. the interaction of stars with the interstellar medium. The energy and momentum injected by the newly formed stars will unbind and disperse the matter around them. How quickly this occurs is key to the SFE. Studying feedback in SSCs is essential to understand how such clusters form and evolve.

The specific role of each stellar feedback mechanism on impeding star formation and driving cloud disruption is a debated topic \citep[see review paper][]{krumholz14}. Several stellar feedback mechanisms have been proposed and quantified analytically. \citet{matzner00} discuss the ability of winds from low mass stars to drive turbulence and balance dissipation. \citet{matzner02} discusses the disruption of clouds by photoionization and expansion of the \ion{H}{ii} region by the thermal gas pressure. More recent studies have included the impact of the radiation pressure on the expansion of the \ion{H}{ii} region and the disruption of the parent molecular clouds \citep{murray10,fall10}. The mechanical energy associated with supernova explosions and winds from massive stars are also potentially able to disrupt clouds. The mechanical energy would be thermalized in shocks producing high pressure hot plasma, which can drive the expansion of the surrounding matter, provided it remains within the cloud. Observations and theoretical work indicate that this may not be as efficient as originally proposed \citep{castor75}, because the plasma is able to flow out through holes in the bubbles very early in the expansion \citep{harper09}. 

Analytical models and numerical simulations have attempted to explain how parent molecular clouds of massive ($\sim$10$^5$ \msun) star clusters are disrupted \citep[e.g.,][]{murray10,dale11,dale12,skinner15}. Analytical models suggest that radiation pressure is the dominant stellar feedback mechanism close to the clusters where the escape velocity is larger than the sound speed of the \ion{H}{ii} gas \citep{krumholz09,murray10}.  But, observational evidence of feedback in individual young ($\lesssim$ 3~Myr) and massive ($\gtrsim$10$^5$~\msun) clusters is still scarce.

SSCs in nearby galaxies are ideal sites to investigate feedback mechanisms of massive clusters.  \citet{lopez11} have used observations of the 30~Doradus star-forming region in the Large Magellanic Cloud, powered by one of the closest SSC \citep[$\gtrsim 5\times10^4$~\msun,][]{andersen09}, to compare the relative values of the pressures associated with radiation, ionized gas and hot plasma, as a function of position in the nebula. They claim that the embedded massive cluster has broken apart its parent cloud within a time scale shorter than the main sequence life time of their most massive stars, thus before SN explosions become significant. Within the shells close to the core of the central R136 SSC, radiation pressure dominates.  In this paper, we carry out a similar study in the Antennae galaxy merger for the most massive and youngest SSC across the overlap region.

The Antennae galaxy (NGC 4038/39) is a nearby \citep[22~Mpc,][]{schweizer08} and well studied merger between two gas-rich spiral galaxies. The region where the two spiral disks interact is known as the overlap region. The molecular gas content is principally constrained to both nuclei and to the overlap region, where it is observed to be fragmented into Super-Giant Molecular Complexes \citep[SGMCs,][]{wilson00}. Interferometric observations showed that SGMCs are clumpy and present different velocity components (\citealp[SMA observations by][]{ueda12}, \citealp[ALMA observations by][]{herrera12} and \citealp{whitmore14}). The Antennae galaxies host a large number of  SSCs \citep{whitmore95,mengel05,whitmore10}, which have masses almost up to $10^7$~\msun. We refer to \citet{adamo15} for a recent discussion of the fraction of star-formation that happens in clusters.  In this study we focus in the overlap region, where the most massive ($\geq 10^5$~\msun), and also the youngest ($<10$~Myr), SSCs are located \citep{mengel05}, which are spatially associated with SGMCs.

The paper is organized as follows. In Section~\ref{sec:obs}, we introduce the datasets used in this paper.  
In Section~\ref{sec:elec}, we present the search in our field-of-views for SSCs associated with molecular and ionized gas on scales of their parent GMCs. We focus on a unique SSC with both molecular and ionized gas emission, suggesting that it may still be embedded in its parent cloud. This SSC, B1, is the most massive SSC in the overlap region. Section~\ref{sec:phys} presents the physical characteristics of SSC~B1 and its parent cloud, and the measurement of their gas emission lines. In Section~\ref{sec:feed}, we describe the physical structure of the matter surrounding SSC~B1. In Section~\ref{sec:radp}, we discuss the role of radiation pressure in disrupting the parent molecular cloud and argue that the bound natal cloud is already disrupted and outflowing gas from this cloud can still be observed. In Section~\ref{sec:diss} we discuss our interpretation of the data. Finally, Section~\ref{sec:con} gives the conclusions of this paper.


\section{Observations}\label{sec:obs}

Our investigation combines  two datasets on the Antennae overlap region. We use observations obtained in February 2011 with the SINFONI imager spectrometer facility on the ESO Very Large Telescope, in the near-IR $K$-band which includes Br$\gamma$ and several H$_2$ rovibrational lines. The spectral resolving power is $R=4000$ at $\lambda=2.2~\mu$m. The SINFONI field-of-view (FOV) is 8\arcsec$\times$8\arcsec\ in size. In \citet{herrera11} and \citet{herrera12} we introduced our four SINFONI pointings across the overlap region, each of them covers a single SGMC. Figure~\ref{fig:sscAntennae} shows the $K$-band emission from the Antennae overlap region, where we mark with dashed boxes our four SINFONI FOVs. In this paper, we use the SINFONI observations of SGMC~4/5. The angular resolution is 0\farcs7$\times$0\farcs6 (full width at half maximum, FWHM). We compute the velocity resolution at different wavelengths by fitting Gaussian curves to the sky-lines (see Table~\ref{tab:fluxssc45}). Details of the data reduction can be found in \citet{herrera12}.

We combine the previous dataset with archive observations done with the ALMA interferometer in the Cycle 0 early science process between July and November 2012, which covered the entire Antennae overlap region \citep[PI: B. Whitmore,][]{whitmore14}. The integration time on source was  3 hours and the number of antennas varied between 14 and 24. These observations done in Band~7 (345~GHz) include the \co\ line emission as well as dust continuum emission. The data was retrieved from the ALMA science archive fully reduced by the ALMA team. We image the data using the CLEAN algorithm in CASA, by defining cleaning boxes enclosing the gas emission at each channel.  The native spectral resolution of the observations corresponds to a channel width of $\sim$0.5 MHz ($\sim$0.5 km s$^ {-1}$). To increase the signal-to-noise ratio, we smooth the channels to a velocity resolution of 10 km s$^{-1}$.  The synthesized beam full width at half maximum (FWHM) is 0\farcs55$\times$0\farcs43, which corresponds to a lineal scale of 59~pc~$\times$~46~pc at the distance of the Antennae. We finally correct the image by the primary beam response. 
We also use archive \coo\ (230~GHz) observations obtained with ALMA in the Band~6 during the science verification process. These observations were presented by \citet{espada12}. Data reduction was performed by the ALMA team. We correct the data cube by the primary beam response. The synthesized beam is 0\farcs86$\times$1\farcs67 (PA=77$^{\circ}$), and the velocity resolution is 20~\kms.


\section{Search for embedded clusters}\label{sec:elec}

In this section, we show how we select a unique SSC across the Antennae overlap region to study the disruption of its parent cloud. We first introduce the SSCs population over the overlap region and then we explain how we use the SINFONI data, which cover four FOVs in the overlap region, to search for compact line emission associated with SSCs.

\begin{figure}[!ht]	
  \centering	
  \includegraphics[width=8cm]{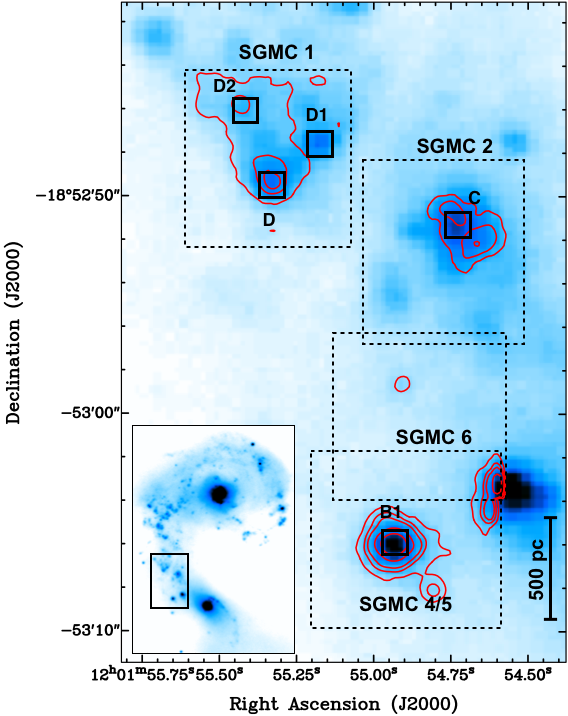}
  \caption{The color image displays the $K$-band emission from the overlap region obtained with the CFHT \citep{herrera11} overlaid with Br$\gamma$ contours from the SINFONI observations. Small black squares highlight the position of the SSCs lying within the four SINFONI FOV, which are marked with dashed boxes. An inserted image of the Antennae galaxies shows, in a black rectangle, the relative position of the overlap region in the merger.}
  \label{fig:sscAntennae}
\end{figure}

All clusters younger than $\sim$10~Myr have significant ionizing flux. For SSCs embedded in their parent cloud, we expect to detect \brg\ line emission from its \hii\ region, which will be compact ($\sim$100~pc) and barely resolved by the arcsecond spatial resolution of near-IR ground-based observations. We also expect to detect H$_2$ line emission from photo dissociated regions (PDRs) \citep[e.g.,][]{habart11}. For older SSCs, the \brg\ emission will be more extended because the ionizing photons will travel into the SGMCs and the more diffuse ISM, over distances larger than the size of their parent cloud. The detection of line emission associated with SSCs in the SINFONI data-cube of the Antennae overlap region is not straightforward because the SGMCs, where SSCs are found, show complex velocity structures and some of them are associated with several SSCs \citep{herrera12}. We face two main difficulties. The first one is to isolate the emission associated with the SSC from that of SGMCs, and the second one is to isolate different sources within SGMCs. We focus on SSCs which are isolated. We have selected a few SSCs previously studied by \citet{gilbert07}, based on single-slit spectroscopic \brg\ observations. \citet{gilbert07} estimated ages and masses from the observed Br$\gamma$ equivalent width and the cluster stellar magnitude,  employing the stellar population synthesis model Starburst99\footnote{http://www.stsci.edu/science/starburst99/docs/default.htm} and adopting a Kroupa initial mass function (IMF). These SSCs are young ($\lesssim$6~Myr) and massive (>$8\times10^5$~\msun), and lie within the four SINFONI fields presented in \citet{herrera12}. These are the clusters D, D1 and D2 in SGMC~1, C in SGMC~2 and B1 in SGMC~4/5, which are highlighted in Figure~\ref{fig:sscAntennae} and listed in Table~\ref{tab:sscs}. We also list in Table~\ref{tab:sscs} the properties of these clusters estimated by \citet{whitmore10} using integrated photometry in the filters UBVIH$_{\alpha}$ complemented with population synthesis models.
For each of these SSCs, we search in the SINFONI data for compact (diameter $\sim$1\arcsec) gas emission associated with the SSCs, to identify which ones are still embedded in their parent clouds. 

\begin{table}[ht]\footnotesize
\begin{center}
\caption{SSCs within the area observed with SINFONI}\label{tab:sscs}
\begin{tabular}{cccccc}
\hline \hline
\noalign{\smallskip}
SSC& Age$^a$ & Mass & Ref.$^b$ &  Flux H$_2$ & Flux \brg\\
ID  &Myr & $10^6~M_{\odot}$ &    &  \multicolumn{2}{c}{erg s$^{-1}$ cm$^{-2}$} \\
\toprule
\noalign{\smallskip}
D	&	3.9		&	1.4		& [1] & <2.9$\times$10$^{-16}$		& 2.1$\pm$0.1$\times$10$^{-15}$ \\
30	& 1.45   & 0.32 & [2] & & \\\cmidrule{1-6}
D1	&	6.1		&	1.6		& [1] & <1.1$\times$10$^{-16}$		& <1.5$\times$10$^{-16}$  \\
 3	& --  & -- & [2] & & \\ \cmidrule{1-6}
D2	& 5.4		&	0.8		& [1] & <1.1$\times$10$^{-16}$		& 6.7$\pm$0.3$\times$10$^{-16}$	\\ \cmidrule{1-6}
C	&	5.7		&	4.1		& [1] & <1.0$\times$10$^{-16}$		& <9.9$\times$10$^{-17}$	\\
28	&4.8  & 1.2 & [2] & & \\ \cmidrule{1-6}
B1$^c$	&	3.5		&	4.2		& [1] & 2.9$\pm$0.2$\times$10$^{-15}$	& 2.2$\pm$0.1$\times$10$^{-14}$  \\
16	&1  & 6.8 & [2] & & \\
\bottomrule
\end{tabular}
\end{center}
Fluxes are not corrected for extinction. \\
$^a$ Errors in ages estimated by \citet{gilbert07} are derived from Starburst99 fits to the Br$\gamma$ EW and are typically below 0.1 Myr. Cluster D2 has a larger error on the Br$\gamma$ EW, the error on the mass is 5.4$^{+0.4}_{-1.4}$ . \\
$^b$ [1] \citet{gilbert07}, [2] Table 8 in \citet{whitmore10}. \\
Masses estimated by \citet{gilbert07} are not corrected by extinction. \\
$^c$ SSC first identified by \citet{whitmore95} as WS80.
\end{table}

For each SSC, we obtain the $K$-band spectrum by circular aperture photometry. The aperture diameter is $\sim$1\farcs3 (140~pc), twice the SINFONI seeing size. We subtract the nearby background measured within an annulus of 0\farcs65 width, the seeing of our SINFONI data. Within the SINFONI spectra, we measure the molecular and ionized emission line fluxes, listed in Table~\ref{tab:sscs}. Upper limits for the fluxes were estimated as the standard deviation of aperture photometry of several positions in the integrated line image. In three of the selected SSCs -- D, D2 and B1 -- we detect \brg. These clusters have ages $\lesssim$5~Myr and masses $\gtrsim$10$^6$~\msun. The young and massive SSC C in SGMC~2, has an ionized nebula larger than the aperture size used to search for compact emission \citep{herrera11}. Other clusters may not be detected in the line emission due to the difficulty of isolating their emission from the background. This is a problem for the less massive clusters. \citet{gilbert07} detected \brg\ emission for all clusters using an aperture size of 2\arcsec\ ($\sim220$ pc), which  includes not only compact \ion{H}{ii} regions but also extended emission. In only one cluster, SSC B1, we detect both H$_2$ and \brg\ compact emission, suggesting that this cluster may still be embedded in its parent molecular cloud.

\begin{figure}[!ht]	
  \centering	
  \includegraphics[width=7.5cm]{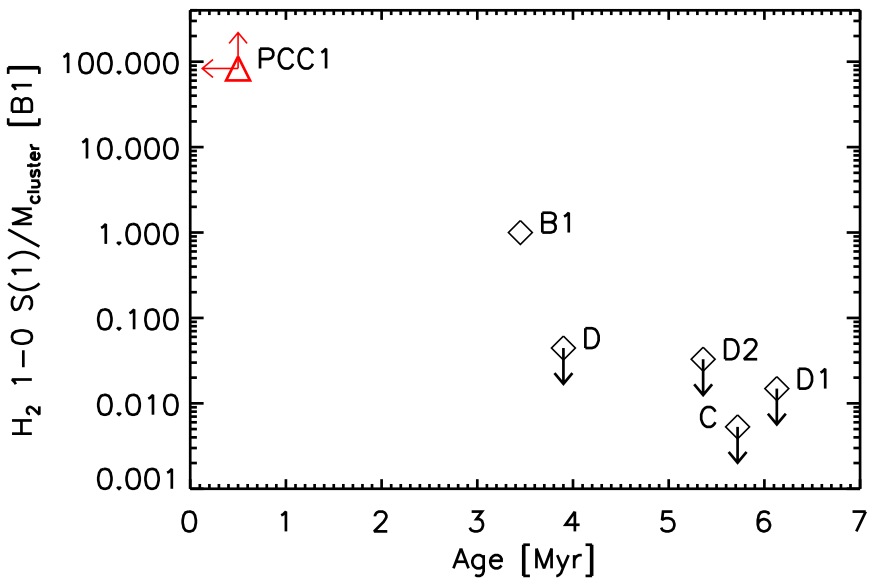}
  \caption{\hhs\ emission over the star cluster mass, for SSCs with different ages and \pcc\  \citep{herrera11}, observed in the SINFONI fields. Since both the masses and \hhs\ fluxes depend on the extinction in the same way, the H$_2$-to-stellar mass ratio is independent of extinction. \pcc\ has a lower limit since it has not yet form massive stars. The H$_2$-to-stellar mass ratio is normalized to the value for B1.}
  \label{fig:h2MassAge}
\end{figure}

\begin{figure}[!h]	
  \centering	
  \includegraphics[width=7.5cm]{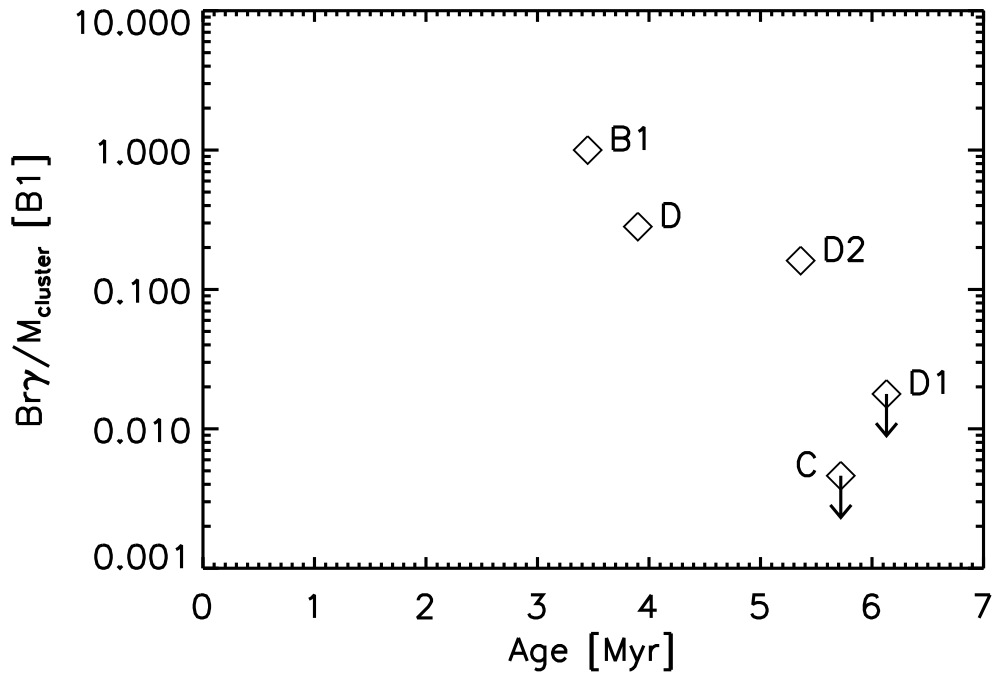}
  \caption{ \brg\ emission over the star cluster mass, for the SSCs in the SINFONI fields. As for H$_2$ (see Fig.~\ref{fig:h2MassAge}), the \brg-to-stellar mass ratio is extinction independent. The ratio is normalized to the value for B1.}
  \label{fig:brgMassAge}
\end{figure}

Figures~\ref{fig:h2MassAge} and \ref{fig:brgMassAge} show the ratio between the line fluxes (both \hhs\ and \brg) and cluster masses versus cluster ages for six and five sources, respectively. The figures include all of the SSCs extracted from \citet{gilbert07}, Fig.~\ref{fig:h2MassAge} also includes the molecular compact source \pcc\ presented in  \citet{herrera11}. The H$_2$-to-stellar mass and \brg-to-stellar mass ratio are normalized to the value for B1. Figure~\ref{fig:h2MassAge} shows that the H$_2$ emission per unit of stellar mass decreases sharply, by more than one order of magnitude for ages of about 3.5 Myr. Only \pcc\ and B1 show clear H$_2$ emission. Sources without an obvious H$_2$ detection are older than B1. This suggests that the disruption of the parent clouds by the newly formed stars occurs within a few Myr (<~5~Myr). Figure~\ref{fig:brgMassAge} shows much scatter. The \brg\ emission does not decrease as fast as the H$_2$ emission.  The comparison between D and D1 in SGMC~1 gives us a clue of the time-scale for the disruption of \hii\ regions in the overlap region. Both clusters have similar masses ($\sim$10$^6$~\msun) but only D, which is 2~Myr younger, is associated with compact ionized gas emission. SSC D1, $\sim$6~Myr old and with a mass of a few times $10^5$~\msun, has already cleared out all or most of its parent cloud. Our findings agree with the upper limit for the parent cloud disruption timescale of < 5~Myr estimated for SSCs in a sample of nearby galaxies and other regions of the Antennae pair \citep{bastian14}.

In the rest of the paper we focus on SSC B1 (also identified as "WS80" in other references), the only SSC across the overlap region associated with bright compact ionized and molecular gas emission. This cluster coincides with a bright emission peak in the mid-IR discovered by ISO, accounting for 15\% of the mid-IR flux from the whole Antennae \citep{mirabel98}. SSC B1 is the brightest compact radio source at 4 and 6~cm in the overlap region \citep{neff00}. It is also one of the brightest sources in the $K$-band and \brg\ emission, in the near-IR observations presented by \citet{gilbert00} and our SINFONI data. \citet{gilbert00} infers that the nebular part of this cluster is a compact, dense \ion{H}{ii} region (n=10$^4$~cm$^{-3}$).  \citet{whitmore02} find that SSC B1 is the intrinsically brightest cluster the Antennae merger. 
The ALMA maps published by \citet[][see their Fig.~1]{herrera12} and by \citet[][their Fig.~10]{whitmore14}  show that it coincides with a CO emission peak. SSC B1 is the more massive and youngest stellar cluster within our SINFONI fields (Table~\ref{tab:sscs}), which are the two fundamental properties to study embedded SSCs without confusion.

  
\section{Physical properties of the SSC and GMC}\label{sec:phys}

SSC B1, in SGMC~4/5, is associated with compact ionized and molecular gas emission. This suggests that B1 may still be embedded in its parent molecular cloud. It is the target we use to study how massive SSCs disperse their parent molecular clouds. In this section, we use the SINFONI and ALMA observations to characterize the physical properties of the cluster and  its parent cloud. 

\subsection{SSC B1}\label{sss:sscb1}

We quantify the number of ionizing photons  $N_{\rm Lyc}$ from the cluster. 
The radio flux of SSC B1 at 4 and 6~cm has a thermal spectral index \citep{neff00}. Both thermal radio emission and hydrogen recombination lines originate within ionized regions and are proportional to $N_{\rm Lyc}$, but the radio emission, unlike the \brg\ emission, is not affected by dust extinction.  From the thermal radio emission, we estimate $N_{\rm Lyc}$ by using Eq.~1 in \citet{romanlopes04}. The radio fluxes at 4 and 6~cm  are listed in Tables~4 and 3 in \citet[][brightest source in their region~\#2]{neff00}, respectively, which gives a mean value of $N_{\rm Lyc}=2.2\pm 0.1\times 10^{53}$~phot.~s$^{-1}$. \citet{gilbert07} report a value $\sim$1.4 lower, estimated from the extinction corrected \brg\ flux. Uncertainties in the estimation of $N_{\rm Lyc}$ come from the unknown fraction of ionising  photons that do not escape the  \ion{H}{ii} region but are absorbed by dust. Such fraction is difficult to quantify. Observations of Galactic  \ion{H}{ii} regions yield photoionizing fraction between 0.3 and 1.0 \citep[see Fig. 10 in][]{draine11} .

The age and stellar mass of  SSC B1 are hard to constrain and discrepancies are found when using different methods. From the $K$-band spectrum of SSC~B1 (Fig.~\ref{fig:ssc45spec}), we can set an upper limit to the age of $\sim$6~Myr due to the absence of CO band-heads at 2.3$\mu$m. This age limit agrees with the ages estimated by \citet{whitmore02} and \citet{whitmore10} of 2~Myr and 1~Myr, respectively, comparing UBVIH$\alpha$ photometry with population synthesis models. It also agrees with the cluster age derived by \citealt{gilbert07} of 3.5~Myr from the equivalent width (EW) of the \brg\ line of 255$\pm$8~\AA, and our own estimation of the \brg\ EW from our SINFONI observations of 295$\pm$24~\AA\ (3.5$\pm$0.7~Myr).  The error bar in our estimation of the age is an statistical value from the EW measurement. Systematical uncertainties from the unknown fraction of ionising photons escaping the  \ion{H}{ii} region are larger.
The stellar mass is estimated to be 6.6$\times$10$^6$ and 6.8$\times$10$^6$\msun\ by \citet{whitmore02} and \citet{whitmore10}, respectively. 

We compute the cluster bolometric luminosity using the Starburst99 models, for a given age and mass of the cluster. Using the age and mass estimated by \citet{whitmore10}, we find L$_{\rm cl}$=5.3$\times 10^9$~\lsun. Note that the bolometric luminosity has a non-linear dependence with age, and the highest value is observed between 2 and 4 Myr. For 3.5~Myr, the age estimated by the Br$\gamma$ EW, the luminosity is $\sim$40\% higher. The mass, age, $N_{\rm Lyc}$ and $L_{\rm cl}$ values of the B1 cluster are listed in Table~\ref{tab:propssc45}. 

In this paper, we choose to use the mass estimated by \citet{whitmore10}, which is obtained at high angular resolution (HST observations), and thus better resolving the stellar emission out of the nebular emission. Additionally, it does not have uncertainties on the photoionisation fraction that escapes the  \ion{H}{ii}  region as our estimation of the stellar mass from the $N_{\rm Lyc}$ value.

\begin{table}[ht]\footnotesize
\begin{center}
\caption{Estimated properties of B1 cluster in SGMC~4/5 and its parent molecular cloud.}\label{tab:propssc45}
\begin{tabular}{llll}
\hline \hline
\noalign{\smallskip}
 $N_{\rm Lyc}$ 					& \multicolumn{3}{l}{$2.2\pm 0.1\times 10^{53}$~phot.~s$^{-1}$} \\
Stellar age$^a$          			& $1-3.5$  Myr  \\
Stellar mass                            		& $6.8\times10^6$  \msun\\
$L_{\rm cl}$        	         		& $5.3\times10^9$ \lsun\\
$A_{\rm K}$ 					& \multicolumn{3}{l}{0.8~mag}\\
Br$\gamma$ emission size 		& \multicolumn{3}{l}{70~pc} \\
Br$\gamma$ Equivalent width		& \multicolumn{3}{l}{295$\pm$28 \AA} \\ 
K-band emission size 			& \multicolumn{3}{l}{66~pc}\\
H$_2$ emission size  				& \multicolumn{3}{l}{$<$210~pc}\\
GMC size$^b$ 					& \multicolumn{3}{l}{104~pc}\\
GMC gas mass $M_{\rm H}$		& \multicolumn{3}{l}{$3.1\pm 0.1\times10^7$~\msun}\\
Mean gas density$^c$ $n_{\rm H}$ 	& \multicolumn{3}{l}{$2.1\pm 0.1 \times 10^3$~cm$^{-3}$}\\
$N_{\rm H}^{c,d}$ 				& \multicolumn{3}{l}{$4.5\pm 0.3 \times 10^{23}$~H~cm$^{-2}$}\\
\hline
\end{tabular}
\end{center}
$^a$ The listed values are those from \citet{whitmore10} and that fitted from Br$\gamma$ EW (see text).\\
$^b$ Size (FWHM) measured from continuum emission at 345~GHz. \\
$^c$ Error bar only includes the error on the cloud mass. \\
$^d$ Column density at the center of the GMC, computed as $N_{\rm H}$=$M_{\rm H}$/Area.
\end{table}

\subsection{Spatial extent of the ionized and molecular gas associated with SSC B1}\label{ss:ext}

We use the \brg\ line to measure the extent of the ionized gas and the continuum emission at 345~GHz for that of the molecular gas. This emission is optically thin and, unlike the near-IR H$_2$ line emission, a tracer of the column density. The size of the ionized emission associated with the cluster is measured by fitting Gaussian curves to the \brg\ emission profiles. The observed size is 0\farcs85$\times$1\farcs02 (FWHM), which after correction by the seeing gives a geometric mean of 0\farcs67 (${\rm FWHM=70}$~pc, listed in Table~\ref{tab:propssc45}). A 2-D Gauss fit of the K-band continuum emission gives a similar observed size of 0\farcs79$\times$1\farcs02 (FWHM), yielding a seeing corrected geometric mean of 0\farcs62 (66 pc).

\begin{figure}[!ht]	
  \centering	
  \includegraphics[width=7.5cm]{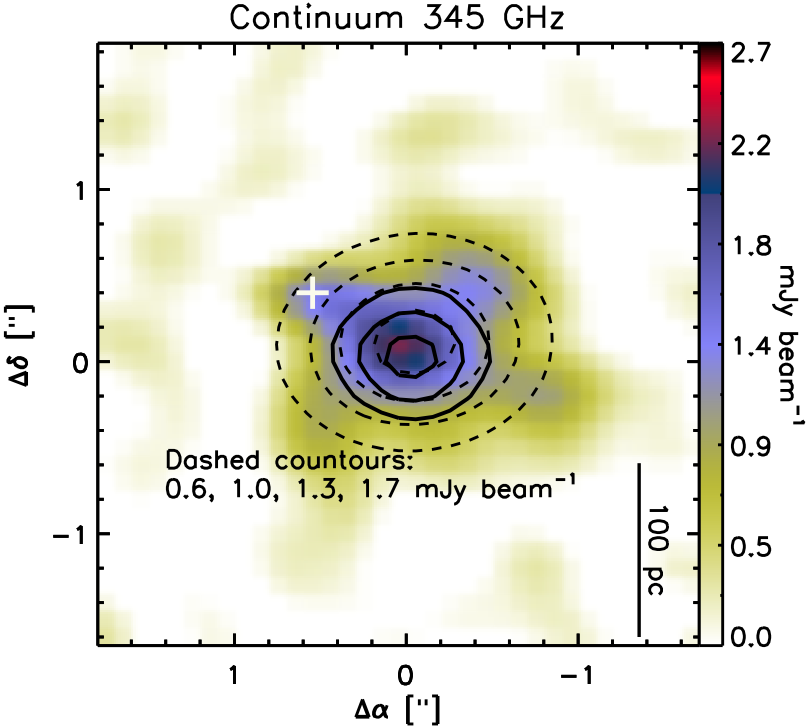}
  \caption{Continuum emission at 345 GHz. Emission from H$_2$O maser was subtracted. The white cross marks the position of the H$_2$O maser.  Solid black lines are the 50\%, 70\% and 90\% of the peak of the \brg\ line emission (see Fig.~\ref{fig:sgmc45}). Dashed black contours correspond to the 2-D Gaussian fit to the continuum emission. Offset positions are relative to $\alpha:$12$^{\rm h}$01$^{\rm m}$54\fs95, $\delta:$ -18\degr53\arcmin05\farcs6 J2000.0. The offset position (0,0) corresponds to the position of the SSC B1, which coincides with the continuum peak emission.}
  \label{fig:cont}
\end{figure}

The size of the molecular cloud is measured by fitting a 2-D Gaussian function to the continuum emission. We first subtract the emission seen to be associated with a water maser\footnote{The absence of a Br$\gamma$, $K$-band continuum or H$_2$ counterpart suggests that this source is an extremely young star forming region.} detected by \citet{brogan10} at the position $\alpha:$12$^{\rm h}$01$^{\rm m}$54\fs9959, $\delta:$ -18\degr53\arcmin05\farcs543, $\sim$0\farcs7 away from SSC B1 (see Fig.~\ref{fig:sgmc45}, peak emission in the continuum at 345 GHz). We measure the size (FWHM) of the emission associated with the water maser to be 0\farcs76 ($\sim$80 pc).  We create a 2-D Gaussian array, centered at the position of the maser,  with a peak emission of half of that seen in the continuum emission at the position of the maser.  This multiplicative factor of 0.5 is found to minimise the dispersion of the residuals after subtracting the water maser contribution and the 2-D Gaussian fit of the GMC emission. Figure~\ref{fig:cont} displays the result of the water maser subtraction, i.e. the continuum emission associated with SSC B1. The black dashed contours correspond to the 2-D Gaussian fit to this emission, black solid contours to the \brg\ emission and the white cross the position of the water maser. The peak of the continuum emission coincides with the position of the SSC~B1, offset (0,0). The  observed source size (FWHM) is 0\farcs96 $\times$ 1\farcs22, which gives a geometrical mean, after beam correction, of 0\farcs98 (${\rm FWHM=104}$~pc). This is the spatial extent of the molecular cloud seen to be associated with SSC B1. It is larger than the \brg\ extent, which suggests that the ionizing emission comes from within the cloud. Gaussian fits to the \hhs\ emission profile yield a size of 2\farcs0$\times$2\farcs1 (FWHM), which after correction by the seeing gives a geometric mean of 2\arcsec (${\rm FWHM=210}$~pc, listed in Table~\ref{tab:propssc45}). This value is larger than the GMC size. This indicates that part of the H$_2$ emission is coming from the SGMC and is not directly associated with SSC~B1. 

\subsection{Intensities of the ionized and molecular gas associated with SSC B1}\label{ss:lines}
In this section, we describe the measurements of the CO and near-IR line fluxes (H$_2$, \brg, \ion{He}{i} and Br$\delta$), listed in Table~\ref{tab:fluxssc45}.

\begin{figure*}[!ht]
  \centering
  \includegraphics[width=15cm]{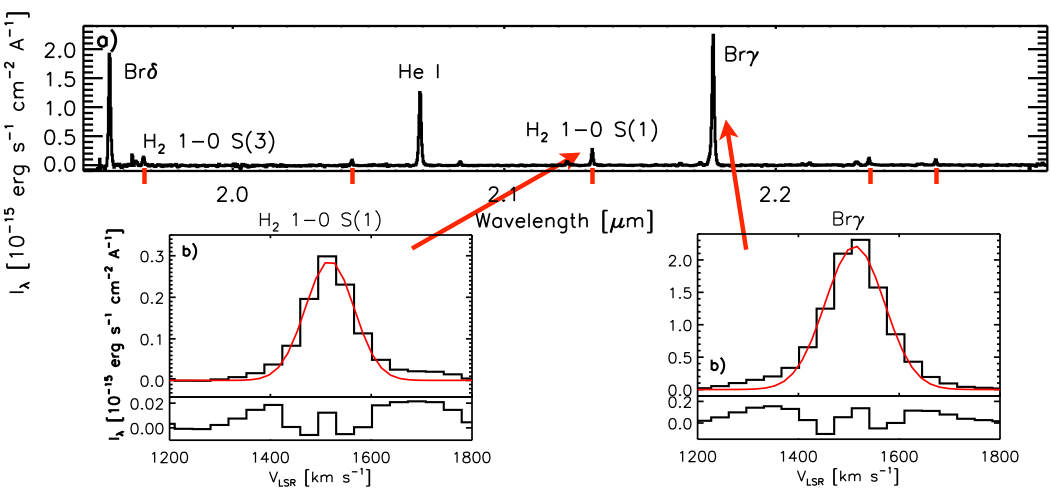}
     \caption{Upper panel displays the $K$-band spectrum of SSC B1, obtained from aperture photometry based on the \brg\ line. Ionized gas lines are the brightest lines across the spectrum. We mark the position of the H$_2$ rovibrational lines with short red lines. Lower panels correspond to a zoom in the \hhs\ and Br$\gamma$ lines. Red lines represent the fitted Gaussian curves to each line, and residuals are plotted in the bottom part. Table~\ref{tab:fluxssc45} lists the parameters of the Gaussian fits.}
        \label{fig:ssc45spec}
  \end{figure*}  
  
The $K$-band spectrum of SSC B1 is presented in Fig.~\ref{fig:ssc45spec}. Several ionized and molecular lines are detected, we highlight the main ones. This spectrum is obtained integrating the flux within an aperture of twice the FWHM observed in \brg\ (1\farcs9 diameter), and subtracting a background estimated within an annulus of 0\farcs65 width (the size of the seeing disk of the SINFONI data). The same figure shows a zoom on the Br$\gamma$ and \hhs\ lines. \brg\ is one order of magnitude brighter than the \hhs\ line.  Since the H$_2$ spatial extent is larger than that of \brg, the H$_2$ fluxes measured from this $K$-band spectrum represent a rough estimate of the H$_2$ emission associated with SSC B1. For a larger aperture of 4\farcs2 diameter,  twice the FWHM of the H$_2$ emission profile, with no background subtraction since it is negligible, we measure a \hhs\ flux four times larger. The aperture size of 4\farcs2 is likely to be dominated by the SGMC (conservative upper limit). Measurements of the line intensities are done by fitting Gaussians to each emission line, and are listed in Table~\ref{tab:fluxssc45}. For most of the lines, we estimate the Gaussian fit from a velocity range limited to the central channels.  Figures~\ref{fig:ssc45spec} a and b show the Gaussian fit to the \hhs\ and \brg\ lines, as well as the residual of these fitting.  Fluxes in Table~\ref{tab:fluxssc45} are not corrected for extinction.  Errors in fluxes correspond to the errors obtained from the Gaussian fit, to which we include as standard measurement errors the quadratic sum of the standard deviation per channel $\sigma_{\rm chan}$ (close to the line emission)  and the 10\% uncertainty of the SINFONI absolute flux. Velocities are given with respect to the LSR with error-bars estimated from the fitting of OH sky lines ($\sim$ 1.5--3 \kms). Table~\ref{tab:fluxssc45} also lists the observed line widths (FWHM) derived from the the Gaussian fits, the spectral resolution for each line and the intrinsic line width (observed value deconvolved by the spectral resolution).

We estimate the extinction in the $K$-band by comparing the $N_{\rm Lyc}$ values obtained with the thermal radio emission (Section~\ref{sss:sscb1}) and that obtained with the \brg\ emission. We compute $N_{\rm Lyc}$ from \brg\ assuming case B recombination, an electron density $n_{\rm e}=10^{4}$~cm$^{-3}$ and an electron temperature of $T_{\rm e}=10^{4}$~K \citep[see Table~6 of ][]{hummer87}.  We estimate an extinction for \brg\ of 0.8 magnitudes (a factor 2), listed in Table~\ref{tab:propssc45}. \citet{gilbert00} reported a slightly higher value of 1.2~mag. Their \brg\ flux, measured from single-slit spectroscopy, is 1.5 times smaller than ours. \citet{whitmore10} reported E(B-V) to be 2.44, and assuming a Draine extinction law, we obtain A$_{K}$=0.9, which agrees with our estimated value.

\begin{table*}[ht]\footnotesize
\begin{center}
\caption{Line parameters measured from SSC B1 in SGMC~4/5}\label{tab:fluxssc45}
\begin{tabular}{ccccccc}
\hline \hline
\noalign{\smallskip}
Line & Rest wavelength& Flux$^{a}$ & V$_{\rm LSR}$ &FWHM$^b$& Resolution &FWHM$^c$\\
& $\mu$m& erg~s$^{-1}$~cm$^{-2}$& \kms\ & \kms\ & \kms\ & \kms\ \\
\hline
\noalign{\smallskip}
Br$\delta^{d}$	    &1.94509		&$1.8\pm0.1\times 10^{-14}$	& $1499\pm3$	& $147\pm5$	& $129\pm6$ 	& $67\pm10$  \\
HeI               & 2.05869 		&$1.4\pm0.1\times 10^{-14}$	& $1521\pm3$	& $143\pm7$	& $106\pm3$ 	& $96\pm10$	\\
HeI               & 2.1127 		& $6.5\pm0.5\times 10^{-16}$	& $1511\pm3$	& $126\pm5$	& $101\pm6$	& $76\pm11$	\\
Br$\gamma$& 2.16612		& $2.4\pm0.2\times 10^{-14}$	& $1512\pm3$	& $137\pm7$	& $87\pm4$	& $106\pm9$	\\

H$_2$ 1$-$0 S(3) &1.95756	& $1.7\pm0.1\times 10^{-15}$	& $1501\pm4$	& $144\pm4$ 	& $129\pm6$	& $64\pm15$ \\
H$_2$ 1$-$0 S(2) &	2.03376	& $9.8\pm0.7\times 10^{-16}$	& $1518\pm3$	& $129\pm5$	& $106\pm3$	& $73\pm10$\\
H$_2$ 1$-$0 S(1) & 2.12183	& $2.5\pm0.2\times 10^{-15}$	& $1518\pm3$	& $113\pm6$ 	& $101\pm6$	& $51\pm17$ \\
H$_2$ 1$-$0 S(0) &	2.22329	& $1.3\pm0.1\times 10^{-15}$	&$1511\pm3$ 	& $118\pm4$ 	& $86\pm6$	& $80\pm9$ \\
H$_2$ 2$-$1 S(1) &2.24772	& $9.6\pm0.8\times 10^{-16}$	&$1515\pm3$	&  $109\pm6$ 	& $86\pm6$	& $67\pm12$  \\
\hline
\noalign{\smallskip}
& GHz & Jy \kms\ & \kms\ & \kms\ & \kms\ & \kms\ \\
\hline
\noalign{\smallskip}
\coo\	$^{e}$ 		 		& 230.538	& $110\pm2$&$1504.4\pm0.2$	& $60.5\pm0.4$	& 20 	&  $57.1\pm0.2$ \\
		    				&		& $80\pm2$	&$1612\pm1$		&  $106\pm2$	& 20	& $104\pm2$ \\
\co\ $^{f}$           			& 345.796& $153\pm 4$	&$1502\pm1$	 	& $59\pm1$ 	& 10	&  $58\pm1$ \\
\co\ $^{g}$	&		& $115\pm6$	& $1594\pm2$ 		& $100\pm5$  	& 10	&  $100\pm5$ \\				
\hline
\end{tabular}
\end{center}
The error in the fluxes for all the lines includes the 10\% of absolute flux uncertainty and rms per channel. Errors in the central velocities include the uncertainty on the OH line identification ($\sim$1.5 -- 3~\kms). \\
$^{a}$ Near-IR fluxes are not corrected for extinction. \\
$^b$ Observed line-widths, without correction for the spectral resolution. \\
$^c$ Line-widths corrected by the spectral resolution. \\
$^{d}$ The Br$\delta$ line is located in a noisy region towards the end of the spectrograph, and shows confusion with night sky lines (OH lines) at 1.945 $\mu$m.\\ 
$^{e}$ Low and high velocity components measured from an aperture of 4\farcs7, twice the source size defined in the CO(2--1) integrated
emission. \\ 
$^{f}$ Low velocity component measured from an aperture of 2\farcs9.\\
$^{g}$ High velocity component measured from an aperture of 4\farcs2.
\end{table*}

\begin{figure*}[!ht]	
  \centering	
  \includegraphics[width=16cm]{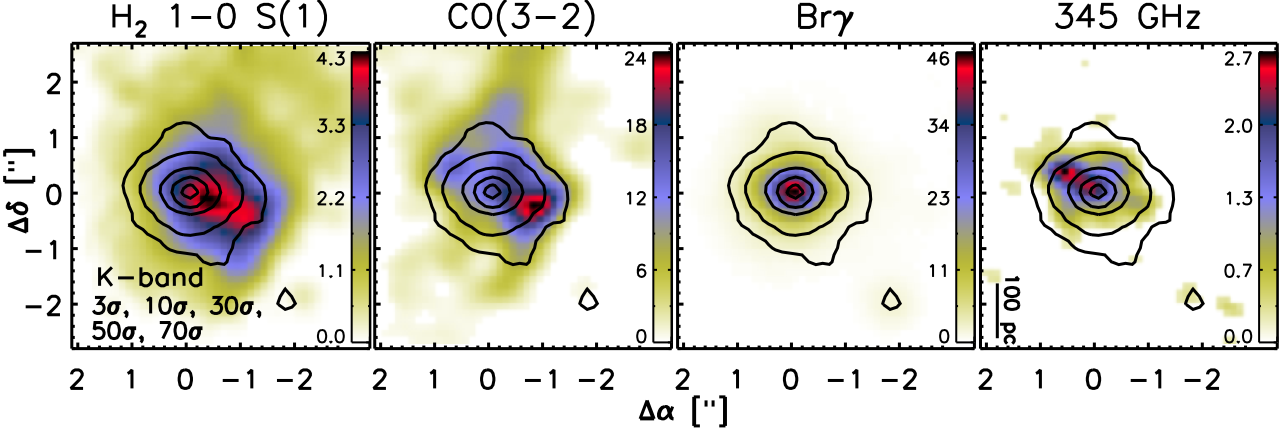}
  \caption{Intensity maps for SGMC~4/5. From the left to right, $K$-band continuum emission contours overlaid on the \hhs\ emission,  \co\ emission,  \brg\ emission and continuum emission at 345~GHz. Flux units of the color bars are $\times 10^{-17}$ erg s$^{-1}$ cm$^{-2}$ for H$_2$ and Br$\gamma$, Jy beam$^{-1}$ for \co, and mJy beam$^{-1}$ for the continuum emission. The 1$\sigma$ emission level of the $K$-band continuum emission is $2.05\times10^{-20}$~erg~s$^{-1}$~cm$^{-2}$. The peak emission of the K-band contours marks the position of the SSC B1 which corresponds to the offset position (0,0). Offset positions are as in Fig.~\ref{fig:cont}.}
  \label{fig:sgmc45}
\end{figure*}

\begin{figure*}	
  \centering	
  \includegraphics[width=18cm]{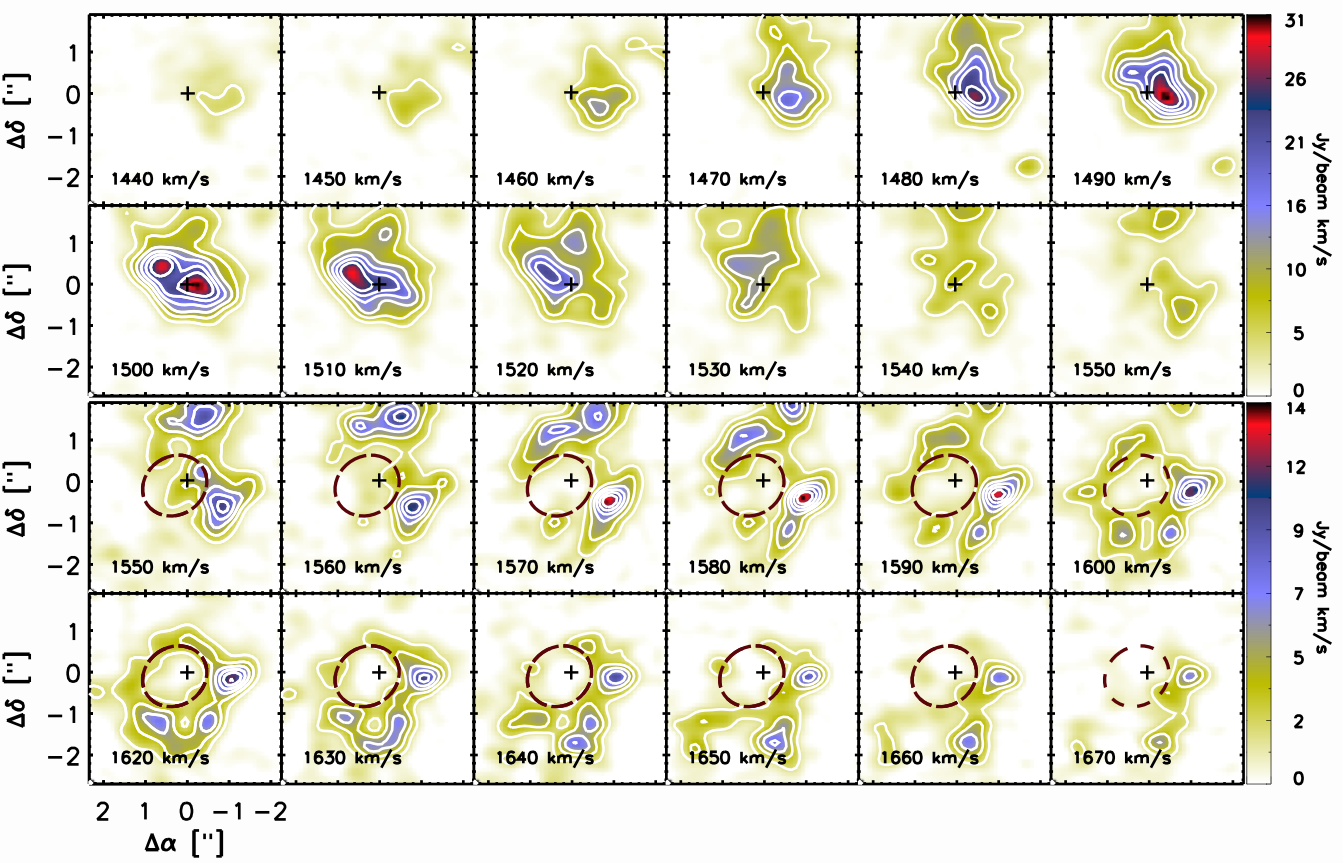}
  \caption{\co\ channel maps of SGMC~4/5.  The position of  SSC~B1 is marked with a cross and corresponds to the (0,0) offset position. {\it Top panel}. We present channel maps from 1440 km s$^{-1}$ to 1550 km s$^{-1}$ by 10 km s$^{-1}$.  Contours start at 10$\sigma$ with a spacing of 10$\sigma$, with $\sigma$=0.39 Jy~beam$^{-1}$~km~s$^{-1}$ measured in line-free channels.  {\it Bottom panel}. CO channel maps from 1550 km s$^{-1}$ to 1670 km s$^{-1}$ by 10 km s$^{-1}$.  Contours start at 5$\sigma$ with a spacing of 5$\sigma$.  Offset positions are as in Fig.~\ref{fig:cont}.}
  \label{fig:chan}
\end{figure*}

\begin{figure}[!ht]	
  \centering
  \includegraphics[width=8.cm]{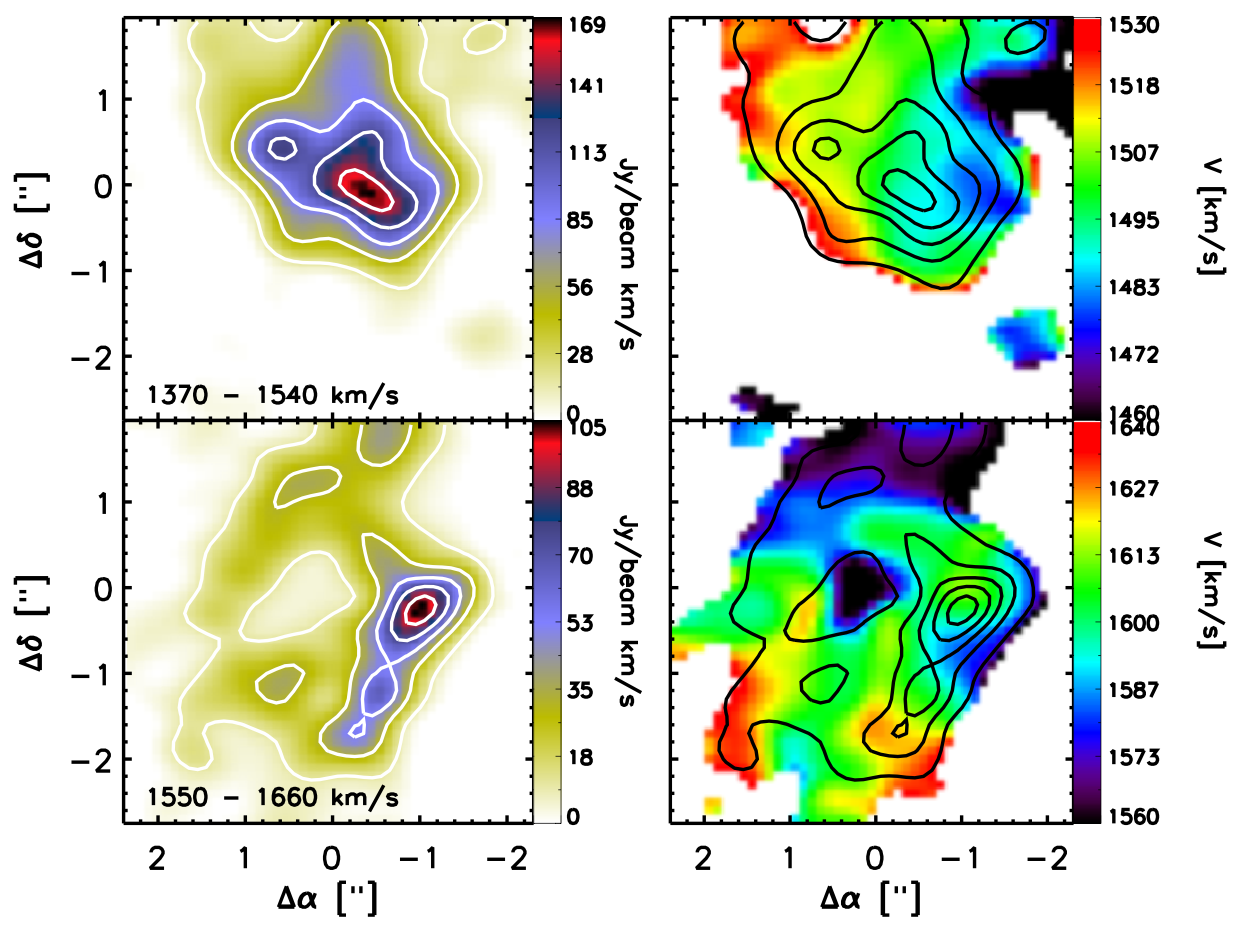}
     \caption{\co\ velocity components in SGMC~4/5. The left panels show the CO emission integrated between 1370 \kms\ and 1540 \kms\ (top) and 1550 \kms\ and 1660 \kms\ (bottom). Contours represent the 10\%, 30\%, 50\%, 70\% and 90\% of the peak emission. Right panels show the first moment of both velocity components. Contours are the same as in left panel. Offset positions are as in Fig.~\ref{fig:cont}.}
        \label{fig:ssc45COvel}
  \end{figure} 
\begin{figure}[!ht]
  \centering
  \includegraphics[width=8.5cm]{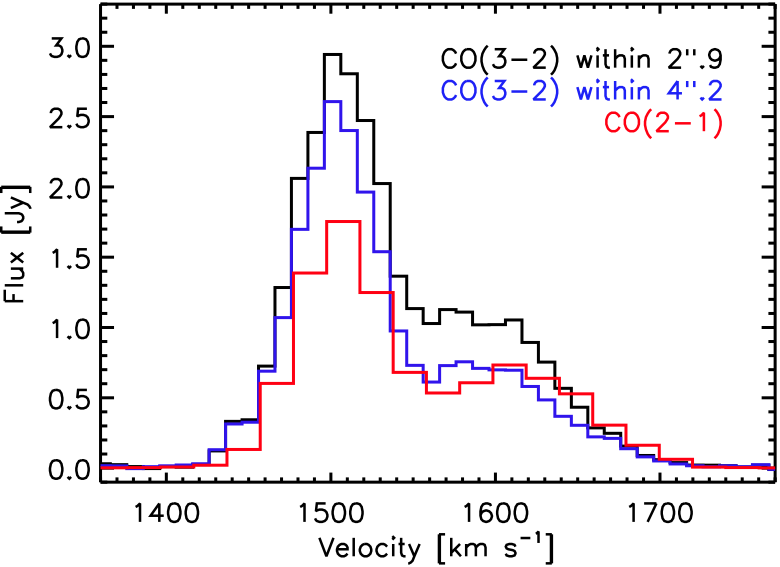}
     \caption{\co\ and \coo\ spectra associated with SSC B1. For the CO J=3$-$2 transition we present the spectra measured with two apertures, 2\farcs9 and 4\farcs2, to characterize the two velocity components observed in the channel maps in Fig.~\ref{fig:chan}.}
        \label{fig:ssc45specCO}
  \end{figure}
  
In Figure~\ref{fig:sgmc45} we have overlaid the $K$-band continuum emission on the \hhs, \co, \brg\ line emission and continuum emission at 345 GHz.  The peak emission of the $K$-band continuum and Br$\gamma$ emission coincides, they define the position of the SSC. The peak in the dust continuum emission at 345 GHz coincides with a water maser discovered by \citet{brogan10}. Figure~\ref{fig:cont} displays the continuum emission with the water maser subtracted, its emission peak coincides with the position of the SSC. The molecular gas, i.e. H$_2$ and CO emission, is more extended than the ionised gas. Neither H$_2$ or CO peaks exactly at the position of the SSC. 
Figure~\ref{fig:chan} presents the \co\ channel maps from 1440 to 1670 \kms.  From these channel maps we can distinguish two spatially and spectrally separated components. Figure~\ref{fig:ssc45COvel} presents the spatial extent (zero moment map) and first moment map of these components. The left top and bottom panels show the \co\ emission integrated over the velocity ranges 1370--1540~\kms\ and 1550--1660~\kms, respectively.  There is a spatial offset between the two velocity components. The low velocity emission is mainly associated with the SSC, while the higher velocity component seems to surround the cluster in a bubble-like shape structure, having the spatial extent of the SGMC. This bubble-like shape was already highlighted by \citet[][see their Fig.~4]{whitmore14}. The first moment maps show that the low velocity component has a velocity gradient while the high velocity component is almost at the same velocity.

 We measure the \co\ line flux of the low velocity component within an aperture diameter of 2\farcs9, twice the observed FWHM size defined in the \co\ emission integrated between 1370 \kms\ and 1540 \kms\ (see top-left figure in  Fig.~\ref{fig:ssc45COvel}), and centered on the peak emission of this component. The resulting spectrum is plotted with a black line in Fig.~\ref{fig:ssc45specCO}. It has a non-Gaussian profile. Line parameters were estimated by fitting 2 Gaussian curves, but only those of the low velocity component are listed in Table~\ref{tab:fluxssc45}. Fig.~\ref{fig:ssc45COvel}  shows that the high velocity component has a larger spatial extent. We measure the \co\ line flux of the high velocity component within an aperture of 4\farcs2 diameter, estimated from the size of the \co\ emission integrated in the high velocity channels. This size coincides with that measured in the \hhs. The result is plotted with a blue line in Fig.~\ref{fig:ssc45specCO}. The parameters listed in Table~\ref{tab:fluxssc45}  for the high velocity \co\ emission component are derived from the fit of this spectrum with 2 Gaussian curves. To quantify the \coo\ emission we used an aperture diameter of 4\farcs7 (red line in Fig.~\ref{fig:ssc45specCO}), size computed from the total \coo\ integrated emission. We fit the emission with 2 Gaussian curves.  Table~\ref{tab:fluxssc45} lists their line parameters. We do not subtract the water maser contribution to the CO emission, as we did for the continuum emission (Section~\ref{ss:ext}), because the velocity structure is very complex making it difficult to separate the contribution of the water maser from the total CO emission. However, this shortcoming does not affect the conclusions of this paper. Systematic uncertainties in the CO-to-H$_2$ conversion factor are much larger than the potential contribution of the maser to the CO emission.
 
    Fig.~\ref{fig:ssc45specCO} displays the \co\ spectra from both apertures mentioned above as well as the \coo\ spectra. Both CO transitions show the two velocity components observed in the channel maps.  From the Gaussian fit we find that the high velocity component is almost twice wider than the low velocity component.  In Table~\ref{tab:vels}, we list the \co-to-\coo\ line ratios for both components. The ratio was measured within an aperture of 4\farcs7 defined by the \coo\ size and after imaging the \co\ observations using the \coo\ synthesized beam as restored beam during the imaging process (CLEAN algorithm in CASA). The intensity ratio between the J=3--2 and 2--1 CO transitions depends on the gas temperature and density as discussed by \citet{schulz07} in the modelling of their single dish observations of the Antennae.  We do not see a significant difference in the excitation of the gas between the low and high velocity components.

 \begin{table}[ht]\footnotesize
\begin{center}
\caption{Parameters of the two CO velocity components.}\label{tab:vels}
\begin{tabular}{lcc}
\hline \hline
\noalign{\smallskip}
    &  Low velocity & High velocity\\
    \hline
    \noalign{\smallskip}
M$_{\rm mol}$							&  $5.1\pm0.2\times10^7$~\msun	& $3.4\pm0.2\times10^7$~\msun \\
    \noalign{\smallskip}
$\dfrac{\rm CO(3-2)}{\rm CO(2-1)}$ 			& $\sim$0.7 					& $\sim$0.6 \\
    \noalign{\smallskip}
$\dfrac{\rm H_2\,2-1\,S(1)}{\rm H_2\,1-0\,S(1)}$ & 0.32$\pm$0.04 				& <0.18 \\
\hline
\end{tabular}
\end{center}
\end{table}

\subsection{Mass of the GMC}\label{ss:massgmc}

We estimate the mass of the GMC, associated with SSC B1, from the continuum emission at 345~GHz. We approximate the spectral dependence on the dust emission with a gray body. We assume that the dust properties in the Antennae overlap region are the same as those in the Galaxy. \citet{scoville14} discuss the use of the Galactic values from Planck observations \citep{planck11a25} to estimate the millimetric masses in local starburst galaxies. They also show that dust temperatures between 15 and 30~K are typically found in starbursts \citep[see also][]{dunne11}, even in extreme cases such as Arp~220, where the dust temperature does not exceed 45 K.  In the Antennae, metallicities of about the solar one are observed towards several SSCs \citep{bastian09}. We assume a typical dust temperature of 25~K and the emissivity per hydrogen measured in the Taurus molecular cloud by \citet{planck11a25}, which is similar to that obtained for starburst galaxies  \citep{scoville14}. Using Eq. 8 in \citet{herrera13}, we derive the mass of the giant molecular cloud of $M_{\rm GMC}= 3.3\pm 0.1\times10^7$~\msun, listed in Table~\ref{tab:propssc45}. The error is estimated as the 1-$\sigma$ standard deviation of the continuum image measured with the same aperture over nearby regions free of sources. Uncertainty on the dust temperature implies an uncertainty of a factor 2  in the mass (if T$_{d}$ is 15~K, the mass will be twice the estimated value, while for T$_{d}=45 K$, the mass will be half the estimated value). The  gas column density $N_{\rm H}=4.5\pm 0.3 \times 10^{23}$~H~cm$^{-2}$ is estimated assuming spherical symmetry. 
 
 The virial mass, assuming a spherical cloud with a density profile $\propto r^{-1}$,  can be written as $M_{\rm vir} = 190\, \Delta v^2\,R$ \citep{maclaren88}. We estimate the virial mass of the low velocity component to be $\sim4\times10^7$\msun. We also estimate the molecular gas mass from the observed \co\ emission. 
We assume a CO-to-H$_2$ conversion factor of $0.6\times 10^{20}$~H$_2$~cm$^{-2}$ (K km s$^{-1}$)$^{-1}$, estimated by \citet{zhu03} for the overlap region at scales of $\sim$1.5~kpc and comparable to the value estimated by \citet{kamenetzky14} using CO modelling at 4.5~kpc scales (Herschel FTS data) in the overlap region. This value is also comparable to the typical value found for starburst galaxies, ULIRGs and towards the Galactic centre \citep[for a detailed discussion about the uncertainties on this factor see][]{bolatto13}.  Other authors have estimated the conversion factor to be similar to the Galactic one \citep[e.g., ][]{wilson03, schulz07}. In this paper we choose to use a starburst-like value, which gives us a good agreement with our other measurements of the molecular gas (see below). 
We also assume a \co-to-CO(1$-$0) ratio of $\sim$0.6, as measured by \citet{schulz07} with single dish data in the overlap region and \citet{ueda12} with interferometric observations at 500 pc scales. Using Eq.~3 from \citet{bolatto13} we estimate $M_{\rm mol}$ to be $5.1\times10^7$~\msun. Using the Galactic CO-to-H$_2$ conversion factor of $2\times 10^{20}$~H$_2$~cm$^{-2}$ (K km s$^{-1}$)$^{-1}$ \citep{solomon87}, the  mass would be more than 3 times higher. The estimated $M_{\rm mol}$  value agrees with the value estimated from the continuum emission and the virial theorem, indicating that the CO-to-H$_2$ conversion factor estimated at 1.5~kpc scales by \citet{zhu03} is also adequate for SGMC~4/5. The measured flux from the high velocity component yields a mass of $3.4\times10^7$~\msun. We list these values in Table~\ref{tab:vels}.


\section{Physical structure of the surrounding matter}\label{sec:feed}

In this section, we describe the physical structure of the molecular gas surrounding SSC~B1 which we associate with the narrow velocity component of the CO spectrum in Fig.~\ref{fig:ssc45specCO} and Table~\ref{tab:fluxssc45}.  We measure the molecular gas pressure and conclude that there is no trapping of IR light within the internal cavity (Section~\ref{ss:press}). In Section~\ref{ss:hot}, we estimate the hot plasma pressure and find that the gas around SSC~B1 is distributed in a inhomogeneous, already broken shell. 

\subsection{Gas pressure in molecular gas}\label{ss:press}

Radiation pressure $P_{\rm rad}$  is the pressure exerted by the stellar radiation over the molecular gas that surrounds the central star cluster. We quantify the  radiation pressure in the molecular gas surrounding SSC~B1. We employ the expression from \citet{murray10} which includes the opacity of the shell,
\begin{equation}\label{eq:prad}
P_{\rm rad} = (1+\langle\tau_{\rm rad}\rangle)\frac{L_{\rm cl}}{4\,\pi\,R_{\rm in}^2\,c},
\end{equation}
where $L_{\rm cl}$ is the luminosity of the cluster, $R_{\rm in}$ is the \hii\ radius and $\langle\tau_{\rm rad}\rangle$ accounts for the trapping of the IR photons\footnote{UV and visible photons emitted by the  cluster are absorbed by the dust grains in the cloud, and re-emitted in the IR.} in the cloud ($\langle\tau_{\rm rad}\rangle\geq0$). The radiation pressure is estimated from the bolometric luminosity emitted by the SSC, {\bf $L_{\rm cl} = 5.3\times 10^{9}$~\lsun} and the \ion{H}{ii} region radius $R_{\rm in}=35$~pc (\brg\ size). We compute
\begin{equation}\label{eq:pradval}
\frac{P_{\rm rad}}{k_{\rm B}}=3.4\times 10^{7}\times (1+\langle\tau_{\rm rad}\rangle)\times\left(\frac{[35~{\rm pc}]}{R_{\rm in}}\right)^{2}{\rm ~K~cm^{-3}}. 
\end{equation}
There are large uncertainties for the age and mass of the cluster (Section~\ref{sss:sscb1}) and the value given in Eq.~\ref{eq:pradval} is a lower limit obtained for the lowest value of $L_{\rm cl}$ in Table~\ref{tab:propssc45}. If we use the $L_{\rm cl}$ estimated for 3.5 Myr and 1.5$\times$10$^7$~\msun, the age and mass computed from the SINFONI data, the radiation pressure is 3 times higher. While the shell  surrounding the cluster is closed and opaque, the IR photons are trapped within the shell, interacting more than once with the dust grains in the shell. In SSC~B1, the moderate value of the effective extinction in the $K$-band (see Table~\ref{tab:propssc45}) preclude large values of $\langle\tau_{\rm rad}\rangle$. 
 
The observational results presented in Section~\ref{sec:phys} allow us to model the physical environment of SSC~B1 (see Appendix~\ref{sec:model} for more details). We model the \ion{H}{ii} region surrounding the SSC as an ionized nebula with dust grains, with incident radiation field of that from the SSC. For different sizes of the internal cavity,  we solve the radiative transfer equation within the \ion{H}{ii} region and measure the outward radiation field and the  pressure exerted at the surface of the PDR. We assume that the gas pressure is set by the radiation pressure and compare our grid of $P_{\rm rad}$ and radiation field values with the outputs of the PDR Meudon code \citep{lebourlot12}, which predicts molecular line intensities from PDRs.  We chose the standard isobaric model with fixed density. These models are detailed in Appendix~\ref{sec:model}.  We find that PDR models with radiation fields in a range of  $\chi$ $\sim$10$^3$--10$^{4}$ times the mean radiation field in the solar neighborhood  and a gas pressure of $3\times10^7-10^8$~K~cm$^{-3}$, are needed to account for the observed H$_2$~1$-$0/2$-$1~S(1) line emission (see Fig.~\ref{fig:CO_PDR}). The value of the radiation field is consistent with that estimated by \citet{gilbert00} from the comparison of PDR models and several near-IR H$_2$ line intensities (obtained with single-slit spectroscopy towards SSC~B1). The estimated gas pressure of $3\times10^7-10^8$~K~cm$^{-3}$ agrees with the measured molecular gas pressure (Eq.~\ref{eq:pradval}), supporting low values of $\langle\tau_{\rm rad}\rangle$. There is no significant trapping of the IR photons within the cavity of the molecular gas surrounding SSC~B1.

\subsection{Gas pressure in the hot gas}\label{ss:hot}

SSC B1 was detected as a compact X-ray source by {\it Chandra} \citep[source \#88 in Table~5 of][]{zezas06}. Its X-ray luminosity, integrated between $0.1-10$~keV, is $L_{\rm X}=1.9\pm 0.2 \times 10^{38}$~erg~s$^{-1}$. This X-ray luminosity  can be used to estimate the electronic density $n_{\rm e}$ and thus the hot gas pressure. 
\begin{equation}
L_{\rm X} =  \epsilon(T) \,\int n_{\rm e}^2\,dl\times A \simeq 4\,\pi\, \epsilon(T)\,n_{\rm e}^2\,R_{\rm in}^3,
\end{equation}
where  $\epsilon(T)=3\times 10^{-23}$~erg~s$^{-1}$~cm$^3$ is the emission coefficient \citep[assuming a temperature of $T=10^7$~K, from figure B.1 in][]{guillard09}, $EM=\int n_{\rm e}^2\,dl\simeq n_{\rm e}^2\times R_{\rm in}$ is the emission measure, and $A=4\,\pi\,R_{\rm in}^2$ is the area where the pressure is exerted. For the observed value of $L_{\rm X}$, we find an electron density of $n_{e}=0.6\times ([35~{\rm pc }]/ R_{\rm in})^{3/2}$~cm$^{-3}$. The pressure of the hot plasma is $P_{\rm hot, X} = 1.9\,n_{e}\,k_{\rm B}\,T$. For a temperature of $T=10^7$~K, the hot plasma pressure is 
\begin{equation}\label{eq:pxray}
\frac{P_{\rm hot,X}}{k_{\rm B}} = 1.2\pm 0.4\times 10^7\times \left(\frac{[35~{\rm pc}]}{R_{\rm in}}\right)^{3/2}~{\rm K~cm}^{-3}. 
\end{equation}
This value is a lower limit because we did not correct the X-ray luminosity for absorption by the shell. Figure~1 in \citet{ryter96} shows that the absorption in the X-ray by the gas is comparable to the extinction in the $K$-band and it is thus a factor of about 2,  increasing $P_{\rm hot,X}$ by a factor 1.4. 

We find that the hot gas pressure is at least 3 times weaker than the pressure of the molecular gas. 
This difference indicates that the hot gas pressure is not dynamically important. The hot plasma produced by shocks driven by stellar winds \citep{castor75} is not trapped within a closed shell. If the mechanical energy from stellar winds over 1 Myr (age of SSC B1) was confined within a closed shell of interstellar matter, then the gas pressure in the hot gas would be more than 50 time higher than that estimated in Eq.~\ref{eq:pxray} (if we use the age and masses estimated from the Br$\gamma$ EW and radio N$_{\rm Lyc}$, then it would be 3 orders of magnitude higher).
This result is consistent with our finding that $\langle\tau_{\rm rad}\rangle$ is negligible. It also implies that the molecular gas interfaces heated by the SSC must be distributed over a range of distances. The estimated molecular radius of $\sim$50~pc (Table~\ref{tab:propssc45}) is a mean value.


\section{Radiation pressure feedback}\label{sec:radp}

In this Section we discuss the role of radiation pressure as a stellar feedback process in SSC~B1.

\subsection{Is the stellar feedback accelerating the gas?}\label{ss:grav}

In Section~\ref{sec:feed}  we show that radiation pressure is the dominant pressure on the molecular gas which is the closest to the cluster, in agreement with previous studies in massive clusters \citep[e.g.,][]{krumholz09,murray10, lopez11}. We compare this  outward pressure with the gravitational attraction exerted by the central cluster and the mass of the cloud itself. For a spherical distribution of the matter, we estimate the inward force per unit surface as,
\begin{eqnarray}\label{eq:grav} 
P_{\rm grav} &=& \frac{1}{4\,\pi\,r^2}\left(\frac{G\,M\,M_{\rm cl}}{r^2}+\frac{G\,M^2}{2~r^2}\right)  \nonumber \\ 
\frac{P_{\rm grav}}{k_{\rm B}}  &=& 1.6\times10^8~\left( \dfrac{52~{\rm pc}}{r}\right)^4{\rm~K~cm}^{-3}, 
\end{eqnarray} 
where $r$ and $M$ are the radius and mass of the GMC, and $M_{\rm cl}$ is the mass of the SSC taken from Table~\ref{tab:propssc45}. Note that the factor 2 in the uncertainty on the GMC mass (see Section~\ref{ss:massgmc}) introduces an uncertainty of a factor 3 in $P_{\rm grav}$. The radiation pressure as estimated in Eq.~\ref{eq:pradval} is weaker than the gravitational force. It is not strong enough to push away the molecular gas associated with SSC~B1. 

In the next section we speculate that most (or at least a significant part) of the pre-cluster gas has already been blown away and that the high velocity component is tracing outflowing molecular gas. From the ALMA observations, \citet{whitmore14} also identifies this component as a super-bubble. Earlier, this gas would have been accelerated by radiation pressure when it was much higher than today.

\subsection{Outflowing gas}\label{ss:outflow}

The momentum of the outflowing gas can be estimated from the molecular mass in the high velocity component of \co\ (taken from Table~\ref{tab:vels}) and the expansion velocity of that gas $v_{\rm exp}$. The latter  value can be estimated using different assumptions that we discuss in Section~\ref{ss:dis}. We find that the expansion velocity of the gas is  $v_{\rm exp}\sim80$~\kms. It is larger than the  escape velocity from the cluster (using the size measured in the $K$-band continuum emission of 66 pc), which is $v_{\rm esc} = \sqrt{2GM_{\rm cl}/R_{\rm cl}}=$42~\kms. Uncertainties on the expansion velocity come from the unknown 3D geometry of the gas and from the mass of cluster. 

To test the hypothesis that this gas was accelerated by radiation pressure over a time scale of $t_{\rm feedback}$, we write,

\begin{equation}\label{eq:mom}
\frac{L_{\rm cl}}{c}\,(1+\tau_{\rm rad})\, t_{\rm feedback} = M_{\rm high}\,v_{\rm exp},
\end{equation}
where
\begin{equation}
t_{\rm feedback}\simeq R_{\rm high}/v_{\rm exp}\simeq 1.2\,{\rm Myr}, \nonumber
\end{equation}
with $R_{\rm high}\sim 100$~pc the mean size of the bubble-like shape structure observed in this component (Fig.~\ref{fig:chan}), and $\tau_{\rm rad}$ refers to the effective opacity of the cloud to radiation (Section~\ref{ss:press}). From Eq.~\ref{eq:mom} we can constrain the value of $\tau_{\rm rad}$  as:
\begin{equation}
(1+\tau_{\rm rad}) = \dfrac{M_{\rm high}\,v_{\rm exp}^2}{\dfrac{L_{\rm cl}}{c}\, R_{\rm high}}\simeq 21\times \left( \dfrac{M_{\rm high}}{3.4\times 10^7\,M_{\odot}}\right)
\end{equation}
This value is higher than that observed today, but within plausible values earlier in the cluster evolution when the parent cloud was not yet disrupted. Indeed, in the models by \citet{murray10}, $\tau_{\rm rad}$ reaches values of several tens at the very beginning of the cluster evolution. Smaller values for $\tau_{\rm rad}$ ($\sim$6) and $v_{\rm exp}$  ($\sim$60~\kms) are found if we use the age estimated from the Br$\gamma$ EW ($L_{\rm cl}$ is higher, see Table~\ref{tab:propssc45}).

We estimate a lower limit for the SFE  within 100~pc from the cluster of 17\%, considering that the high velocity component may include not only natal gas but surrounding gas that did not participate in the cluster formation. We estimate the outflow rate of the parent molecular cloud to be $\dot{\rm M}_{\rm outflow} = M_{\rm high}/t_{\rm feedback} \simeq 30$~\msun/yr.
  
\begin{figure}[!ht]
  \centering
  \includegraphics[width=9cm]{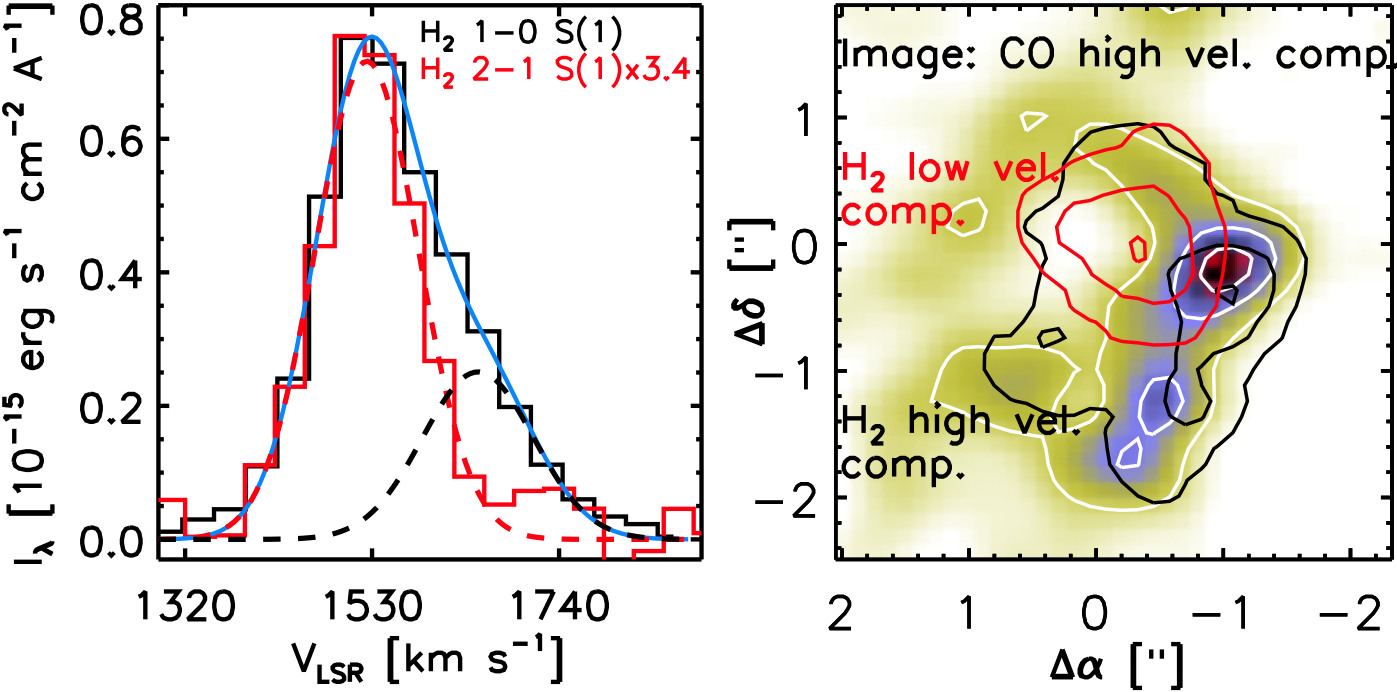}
     \caption{H$_2$~2$-$1/1$-$0 S(1) ratio in  SGMC~4/5. The left panel shows the spectra for the two near-IR H$_2$ lines, \hhs\  in solid-black and H$_2$~2$-$1~S(1) in solid-red line scaled by a factor of 3.4. The dashed-red line is the Gaussian fit for the H$_2$~2$-$1~S(1) line, while the blue-solid line is the fit for the \hhs\ emission that corresponds to the addition of 2 Gaussian curves (the red and black dashed lines). The color image and white contours in the right panel show the integrated intensity of the high velocity component observed in \co. Red and black contours are the H$_2$ emission integrated in 1300--1512 \kms\ (low velocity component) and 1580--1760 \kms\ (high velocity component), respectively. Offset positions are as in Fig.~\ref{fig:cont}.}
        \label{fig:h2_ratios}
  \end{figure}

The broad velocity component is  also observed in the H$_2$~1$-$0~S(1) emission. The left panel in Fig.~\ref{fig:h2_ratios} presents the spectra for the v=1$-$0 and 2$-$1 S(1) H$_2$ line emission integrated within an aperture of 4\farcs2 (Section~\ref{ss:ext}). The v$=$2$-$1 line can be fitted with a single Gaussian curve (red-dashed line in Fig.~\ref{fig:h2_ratios}) while v$=$1$-$0 requires two Gaussian components (blue line in Fig.~\ref{fig:h2_ratios}). We fit the v=2$-$1 line and use the estimated center velocity as a fixed parameter to fit the low velocity component of the  v=1$-$0 emission. The right panel of Fig.~\ref{fig:h2_ratios} shows in red and black contours the low and high velocity components observed in v=1$-$0, respectively, overlaid on the \co\ high velocity component. The extent of the high velocity components seen in \co\ and \hhs\ are similar. Comparing Fig.~\ref{fig:h2_ratios} with Fig.~\ref{fig:ssc45COvel} we find that the extent of the low velocity components seen in \co\ and \hhs\ are also similar.

The  R$_{S(1)}$$=$H$_2$ 2$-$1/1$-$0 S(1) ratio is commonly used to disentangle between H$_2$ gas heated by collisions or UV radiation.
On the one hand, in shocks excitation is collisional and only low-$\nu$ H$_2$ levels are populated because the temperature of molecular gas is at most a few 1000~K;  for higher temperatures the H$_2$ molecule is destroyed in collisions.  On the other hand, H$_2$ excitation through UV pumping and fluorescence populates both high- and low-$\nu$ states \citep[e.g., ][]{herrera11}. We find R$_{S(1)}=0.32\pm 0.04$ for the low velocity component, and estimate an upper limit of 0.18 for the high velocity component using the ratio between the 3$\sigma$ emission for v=2--1 ($\sigma$ measured in the line-free channels) and the peak emission of the high velocity component seen in v=1--0. These values are listed in Table~\ref{tab:vels}. The former ratio can be accounted for by UV heating of the gas in PDRs (see Section~\ref{ss:press} and Appendix A). The latter ratio can be observed towards PDRs with high radiation fields ($>10^4$) and  gas pressure ($\simeq10^8$~K~cm$^{-3}$) (see Fig.~A.1). However, since the high velocity component is extended (a few 100 pc) and there is no evidence for extended massive stellar population, we favor shocks as main gas heating mechanism. Values in the range 0.1--0.2 are reproduced in J- and C-shock models \citep{kristensenphd} for gas with densities in the range 10$^4$--10$^6$~cm$^{-3}$ and shock velocities from 15 to 50~\kms. This supports the idea that the high velocity component is tracing outflowing gas. This interpretation indicates that the 2 velocity components have different excitation mechanisms.

\begin{figure}[!ht]
  \centering
  \includegraphics[width=9cm]{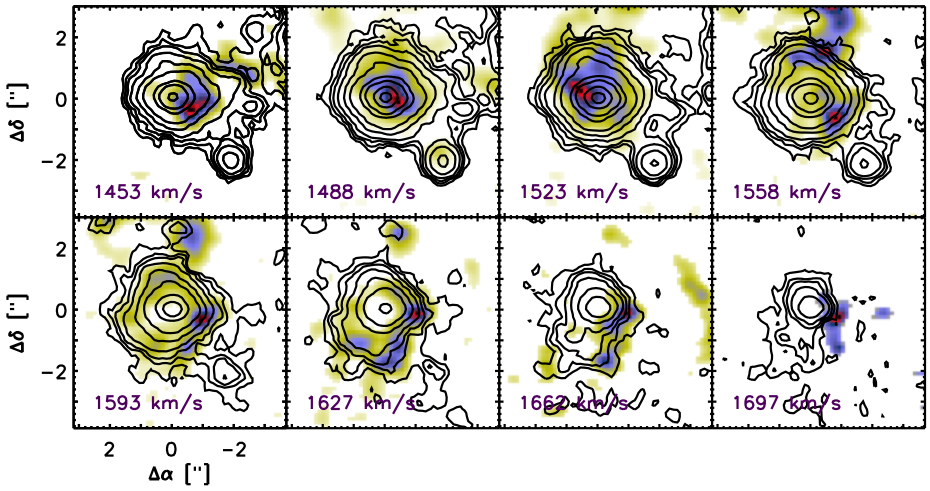}
     \caption{Contours are the channel maps of the ionised gas traced by \brg\ overlaid on the \co\ emission interpolated to the velocity of the SINFONI observations.}
        \label{fig:brgchan}
  \end{figure}

Evidence of high velocity ionized gas can be observed in the \brg\ line emission. Fig.~\ref{fig:brgchan} presents the channel maps of the \brg\ line emission overlaid on the \co\ line emission. At the velocities of the low velocity component ($<~1600$~\kms), the \brg\ emission is symmetrical. An excess of ionized gas emission, where the \co\ has a low column density, is observed at velocities of 1600$\sim$1700~\kms, that we interpret as an evidence of escaping ionised gas.

\section{Possible scenario}\label{sec:diss}

In our interpretation, it is too late to witness the first stages of the disruption of the parent molecular cloud by stellar feedback, when the parent cloud was still a bound cloud. Nevertheless, the high velocity component may be tracing the outflowing gas from this cloud. In the following, we discuss this interpretation and its caveats.

\subsection{Disruption of the parent cloud and outflowing gas}\label{ss:dis}
We propose that radiation pressure was the main mechanism disrupting the parent molecular cloud over a short timescale.  At the beginning of the SSC evolution, radiation pressure was much higher  than today due to the high density and opacity of the parent cloud. The IR photons from the dust heated by the UV photons from the cluster were highly trapped in the cloud implying a high $\tau_{\rm rad}$ value and radiation pressure (see Eq.~\ref{eq:prad}). The momentum transfer rapidly accelerates the surrounding gas to velocities sufficiently high to escape from the cluster. \citet{murray10} modelled this disruption process for a source with similar characteristics (i.e. GMC mass and size, and star cluster mass), finding that the disruption  occurs in less than 1~Myr.

Within this scenario, the high velocity component is tracing  outflowing gas from the parent cloud. The expansion time scale of 1.2 Myr, computed in Section~\ref{ss:outflow}, agrees with the model value from  \citet{murray10}. It is comparable to the age of the cluster (1--3.5 Myr), which makes this scenario of cloud disruption plausible. The bottom panel of Fig.~\ref{fig:chan} shows the channel maps of this component, which surrounds SSC~B1 in a bubble-like shape structure with an inner radius of about 1\arcsec\ ($\sim$100~pc). Similar molecular gas morphology and kinematics are observed in the starburst galaxies M82 and NGC~253 \citep{weiss99,sakamoto06}, where the data have been interpreted as evidence of expanding  molecular super-bubbles. Evidence of expanding bubbles has also been observed for ionised gas \citep[e.g., in M83, ][]{hollyhead15}.

\begin{figure}[h]
  \centering
  \includegraphics[width=8cm]{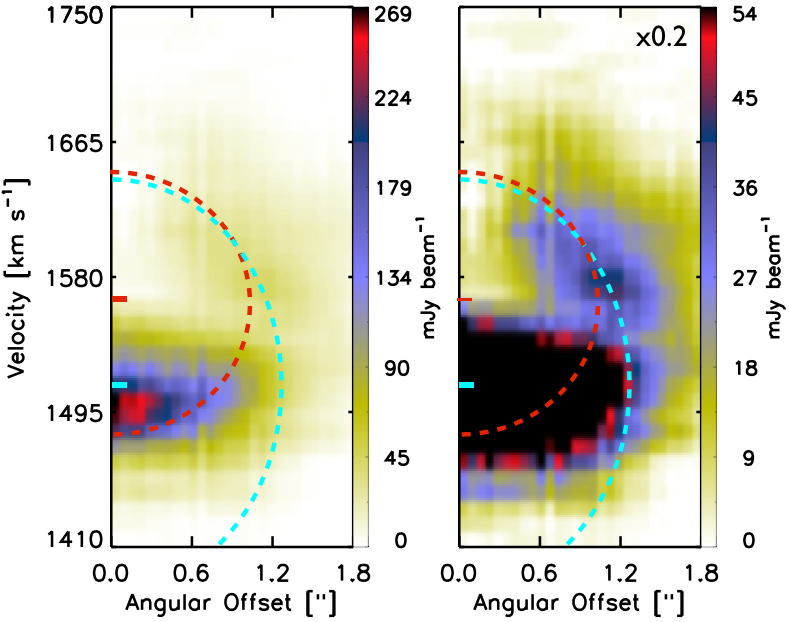}
     \caption{Azimuthally averaged \co\ emission, around the position of SSC B1 up to a maximum radius of 1\farcs8. Left panel shows the total emission and right panel shows the emission clipped up to a maximum value of 20\% of the peak emission. The semi-ellipse in dashed-red line highlight the expanding shell assuming a central velocity of $\sim$1565 \kms\ marked with a red tick on the velocity axis, while for the dashed-cyan line the central velocity is equal to that of the \brg\ emission.}
        \label{fig:shell}
  \end{figure}

In Fig.~\ref{fig:shell}, we show the \co\ emission averaged within concentric annuli centred on SSC~B1 up to a distance of 1\farcs8 from the cluster (azimuthally averaged \co\ emission). For a closed, spherical and homogeneous expanding shell, the azimuthally averaged gas emission should show a semi-ellipse with constant flux. This is not the case for the super-bubble in SSC~B1. The central velocity of the cluster is hard to determine. We can approximate this velocity in two ways. 
First, we assume that the Br$\gamma$ emission is symmetrical (no obscuration by dust) thus use the Br$\gamma$ velocity as a proxy for the cluster velocity. In this case, we approximate the velocity expansion of the bubble as the difference between the Br$\gamma$ velocity and the velocity of the high velocity component, $\sim$80~\kms. Figure~\ref{fig:shell}  shows a cyan dashed-line semi-ellipse to illustrate this scenario. The size of the ellipse was chosen to fit the CO high velocity emission. In this scenario, the CO data do not show any evidence of blue-shifted gas and we have to assume that the CO shell is not symmetric.
Second, we estimate the velocity of the cluster fitting the CO azimuthally averaged emission. The red dashed-line in Fig.~\ref{fig:shell} shows the best fit for the  CO emission (low and high velocity components), with a central velocity of $\sim$1565~\kms, which would be tracing the expanding bubble. In this scenario, the central velocity of the ellipse is about 50 \kms\ larger than that of the compact \brg\ emission seen towards the SSC. This difference with the \brg\ velocity may imply that the red-shifted \brg\ gas is obscured and we only observe the gas that is the closest to us. An expanding ionised bubble with asymmetrical emission has been observed towards a young (6 Myr) and massive ($10^7$~\msun) SSC in a blue compact galaxy \citep{ostlin07}.

\subsection{The molecular gas surrounding SSC B1}

If the parent cloud was disrupted early in the cluster evolution, why do we see a CO emission peak coincident in velocity and position with the cluster? 
Within our scenario, we speculate that the low velocity component traces gas that was near to the SSC when it formed but not part of its parent cloud. It may also include clumps which have migrated from the SGMC environment to the cluster neighbourhood. This gas would have been too far from the SSC to be accelerated by radiation pressure. This possibility is exemplified by the 30 Doradus star-forming region. The gas surrounding the central massive cluster R136 is distributed in an irregular shell, where some molecular clouds that did not participate on the formation of the R136 cluster are observed  very close to it \citep[$\lesssim$ 10 pc, ][]{rubio09}. 

We expect the low velocity molecular gas to be dispersed by the action of stellar winds and future supernova explosions. Stellar winds from massive clusters can erode the surrounding gas and push away the clump material \citep{rogers13}. 

In the scenario that the gas traced by the low velocity component is indeed the parent molecular cloud, we do not really understand how the cluster will expel and disrupt its surrounding matter. This component is still very massive, comparable with the mass of the cluster, and even more it is self-bound. We need observations at higher resolution to resolve this component.
 
\subsection{Star Formation Efficiency}

Is SSC B1 a bound cluster which will evolve as a globular cluster? 
As described in the introduction, a SFE lower limit of 30\% is needed to create a bound stellar cluster. Within our scenario, for a region of $\sim$100~pc surrounding SSC~B1, we estimate a lower limit for the SFE of  17\%. In the extreme, unlikely case that both the low and high CO velocity components were part of the parent molecular cloud, we estimate SFE$>$10\%. If we use the Galactic CO conversion factor, then the SFE decreases to $\sim$5\%. The estimated value is higher than what is measured in Galactic GMCs, but not high enough to ensure the formation of a bound star cluster. To reach the theoretical SFE limit, at least 30\% of the high velocity component should not be natal gas but, for example, surrounding gas that was swept-away by stellar feedback. This is possible for SSC B1, which is located in a complex interacting region rich in molecular gas.  However, as discussed in the introduction, hydrodynamical simulations indicate that even if the global SFE is low, locally the SFE can reach values large enough to form a bound cluster \citep{kruijssen12}. 

Further observations are needed to quantify and test our proposed scenario. Line observations at high resolution of the stellar emission of the cluster would help settle the issues related to the cluster velocity discussed in \S~\ref{ss:dis}. Higher resolution observations will reveal the spatial distribution of the gas very close to the cluster, the low velocity component. With the existing data we cannot resolve structures smaller than 50~pc  and thus we cannot resolve and characterize the clumpy structure of this source.  More sensitive observations will enable us to better trace the morphology of the high velocity component. Also, by looking for high velocity wings, we can elucidate if the molecular gas is been carried out in a wind together with the hot gas. This can now be done with ALMA which offers the required angular resolution and sensitivity to address these questions.

\section{Conclusions}\label{sec:con}
In this paper, we investigate the early evolution of SSC B1 located in SGMC~4/5 in the Antennae overlap region, specifically the impact of the stellar feedback mechanisms on the surrounding matter.

We search for associations between SSCs and compact \ion{H}{ii} and H$_2$ line emission in four FOVs observed with SINFONI/VLT across the Antennae overlap region. We focus on isolated SSCs previously studied by \citet{gilbert07}. Among our sample, there is only one SSC with both \ion{H}{ii} and H$_2$ emission, SSC~B1 in SGMC~4/5. This is the only SSC younger than 5~Myr.  This is probably not a complete sample since it is difficult to isolate the molecular and ionized emission associated with SSCs. Still, our sample shows that the association between SSCs and their parent molecular clouds do not last much more than 5~Myr, similar to what is found in a sample of SSCs in nearby galaxies and other regions of the Antennae pair \citep{bastian14}.

CO emission in SSC B1 presents two velocity components, a compact, low velocity one at the position of the SSC, and an extended, high velocity one distributed in a bubble-like shape structure around the cluster. The virial and millimetric masses of the gas associated with SSC~B1, low velocity component, are comparable. The comparison with its CO luminosity yields a CO-to-H$_2$ factor consistent with that found at larger scales (1.5 kpc) and in other starburst galaxies. 
The detection of molecular gas associated with SSC~B1 suggests that the cluster may still be embedded in its parent molecular cloud.  Although the actual velocity of the cluster is still unknown and thus it is not clear if the cluster is embedded in this cloud, this source is the only one in the overlap region that could tell us how the molecular gas is being pushed away from the cluster.

We analysed the physical properties of SSC B1 and the molecular and ionised gas surrounding it. A model of the physical environment of the \ion{H}{ii} region and the PDR surrounding SSC B1 show that the observed H$_2$~2$-$1/1$-$0~S(1) intensity line ratio is accounted for by a gas pressure of $3\times10^7-10^8$ K cm$^{-3}$, which agrees with the estimated radiation pressure from the cluster.  This finding implies that there is no significant trapping of the IR photons within the cavity of the molecular gas surrounding the cluster. X-ray observations show that the plasma produced by shocks driven by stellar winds is not confined within a closed shell of cold gas. The plasma pressure is smaller than the radiation pressure. However, radiation pressure is weaker than the gravitational force from the cluster and the cloud itself. It is not strong enough to push away the observed molecular gas.

We propose that radiation pressure was early enhanced and was the main mechanism driving the disruption of the parent molecular cloud in  $\sim$1 Myr. The high velocity component would be tracing outflowing molecular gas from this disruption. The association of this component with shock heated H$_2$ gas, and the presence of escaping high  Br$\gamma$ gas, are supporting evidence. The low velocity component may be gas that was near to the SSC when it formed but not part of its parent cloud or clumps that migrated from the SGMC environment. This component would be disrupted by the action of stellar winds and supernova explosions. 
The existing data do not allow us to conclude whether SSC B1 is gravitationally bound and will evolve as a globular cluster. The lower limit for the SFE of 17\% estimated in a region of 100~pc size from the cluster is slightly smaller than the theoretical limit for bound cluster formation. 

Further observations are needed to quantify and test our proposed scenario. Additional observations of line emission from the cluster would allow us to directly measure the velocity of the cluster. Higher spatial and spectral resolution observations are needed to probe the clumpy structure of the closest component to the cluster.

\begin{acknowledgements}
We would like to thank the anonymous referee for the detailed and constructive report that helped us to improve our manuscript. The Atacama Large Millimeter/submillimeter Array (ALMA), an international astronomy facility, is a partnership of Europe, North America and East Asia in
cooperation with the Republic of Chile. This paper makes use of the following
ALMA Science Verification data: ADS/JAO.ALMA\#2011.0.00003.SV and Early Science data: ADS/JAO.ALMA\#2011.0.00876.S.

\end{acknowledgements}

\bibliographystyle{aa}
\bibliography{papers}

\Online

\begin{appendix} 

\section{Modeling the physical environment of SSC B1}\label{sec:model}
 In this section, we introduce a simple geometrical model that we use to quantify the action of the stellar feedback  in the parent cloud of SSC B1. The geometrical model is combined with a PDR model to quantify the H$_2$  line emission.  We introduce the model in two parts, we model the \hii\ region and then the PDR. This 1-D model describes a segment of the shell. It does not assume that the cluster is fully embedded in a spherical shell. The matter around the SSC is likely to be clumpy, with some radiation escaping in some directions. This is in agreement with the fact that the mean extinction of the shell is moderate (a factor 2, \S~\ref{sss:sscb1}). The \hii\ region is directly receiving the UV radiation from the SSC and the incident radiation on the PDR is the output radiation from the \hii\ gas layer. The internal radius of the \hii\ cavity $R_{\rm in}$ is a parameter of the model, the external radius is the GMC radius observed in the continuum emission, $R_{\rm out}=52$~pc. The stellar radiation was chosen to be that from the Starburst99 models, for an age of 3~Myr scaled by the cluster mass. This SED was used for the \hii\ region modeling as the ionizing radiation.

\subsection{\ion{H}{ii} Region}
We model the \hii\ region as an ionized hydrogen nebula with dust grains. We compute the radiative transfer equation for this nebula in the plane-parallel approximation. 
In this nebula, the radiative transfer will be defined by the absorption produced by the ionization of the hydrogen atoms and light absorption and scattering by dust grains,

\begin{equation}
\dfrac{dI(\lambda)}{ds} = -n_{\rm H}\sigma_{\rm H}(\lambda)\,I(\lambda)-n_{\rm d}\sigma_{\rm d}(\lambda)\,I(\lambda) \label{eq:trans},
\end{equation} 
where $s$ is the position in the nebula from $R_{\rm in}$ at which the absorption occurs, $n_{\rm H}$ and $n_{\rm d}$, and $\sigma_{\rm H}$ and $\sigma_{\rm d}$ are the hydrogen and dust densities and cross sections, respectively. $\sigma_{\rm H}$ is zero for non ionizing photons.  We assume that the nebula extent is small compared to $R_{\rm in}$. The ionization rate ($N_{\rm ion}=n_{H}\bar{\sigma}\,S$) equals the recombination rate ($N_{\rm rec}=n_{\rm e}\,n_{+}\,\alpha_{2}$), where $\bar{\sigma}$ is the mean photoionization cross section, $S=S_{0}\,e^{-\tau}$ is the number of ionizing photons per unit time and unit area, $n_{\rm e}$ is the electron density, $n_{+}$ is the density of protons and $\alpha_{2}$ is the recombination coefficient. Then, we can re-write Equation~\ref{eq:trans} as:

\begin{equation}
\dfrac{d\,e^{-\tau_{\lambda}}}{ds} = -\dfrac{\alpha_{2}}{S_{0}}n^{2}-n\,x_{d}\,\sigma_{d}(\lambda)\,e^{-\tau_{\lambda}}, \label{eq:y}
\end{equation}
where $n$ is the gas density and $x_{d}$ is the dust-to-gas mass ratio. We assume that all the protons and electrons come from the same
atoms ($n=n_{\rm e}=n_{+}$). The dust cross sections are taken from the DUSTEM model \citep{dustem90,compiegne10}. We include the extinction cross sections of the very small grains and big grains (there is no observational evidence for PAHs in \ion{H}{ii} regions, i.e. \citealt{kassis06}).

We solve the radiative transfer equation (Eq.~\ref{eq:y}) for different values of incident radiation on the \hii\ region, depending on the value of the unknown quantity $R_{\rm in}$, which we vary from 30 to 100 by spacings of 10~pc. This range is distributed around the GMC radius ($\sim$50~pc), from the radius of the \brg\ line emission to that of H$_2$.  We compute the output flux in units of the Mathis radiation field \citep{mathis83}, integrating the emission between 91.2 to 200~nm. Table~\ref{tab:hii} lists the radiation field for the different internal radii of the cavity. We also compute the radiation pressure exerted at the surface of the PDR from photons absorbed or scattered across the layer of the \hii\ gas,
\begin{equation}
P_{\rm rad}({\rm HII}) = \frac{L_{\rm HII}}{4\,\pi\,R_{\rm in}^2\,c}.
\end{equation}
where the luminosity $L_{\rm HII}$ is the luminosity incident onto the \hii\ region minus the output luminosity. $ P_{\rm rad}({\rm HII})$ is the pressure produced by the ionizing photons and the non-ionizing photons which are absorbed by dust and gas in the \hii\ layer. This is the minimum value of the pressure at the surface of the PDR. 
Table~\ref{tab:hii} lists the estimated values of $P_{\rm rad}$.

\begin{table}[ht]\footnotesize
\begin{center}
\caption[\hii\ region parameters: $\chi$ and $P_{\rm rad}/k_{\rm B}$]{Radiation fields and gas pressures exerted at the surface of the PDR, for different values of the internal cavity of the \hii\ region $R_{\rm in}$.}\label{tab:hii}
\begin{tabular}{ccc}
\hline \hline
\noalign{\smallskip}
$R_{\rm in}$	& $\chi$	& $P_{\rm rad}/k_{\rm B}$\\
pc		& Mathis	&  K~cm$^{-3}$ \\
\hline
\noalign{\smallskip}
30		& 1.6$\times 10^4$ & 1.5$\times 10^8$ \\
40		& 1.1$\times 10^4$ & 8.6$\times 10^7$ \\
50		& 7.5$\times 10^3$ & 5.5$\times 10^7$ \\
60 		& 5.5$\times 10^3$ & 3.8$\times 10^7$ \\
70		& 4.2$\times 10^3$ & 2.8$\times 10^7$ \\
80		& 3.3$\times 10^3$ & 2.1$\times 10^7$ \\
90		& 2.7$\times 10^3$ & 1.7$\times 10^7$ \\
100		& 2.2$\times 10^3$ & 1.4$\times 10^7$ \\
\hline
\end{tabular}
\end{center}
\end{table}

\subsection{PDR}\label{ss:model}

To describe the PDR, we use the online tables of outputs from the PDR Meudon Code\footnote{http://pdr.obspm.fr/PDRcode.html} \citep{lebourlot12}. We chose to use results from isobaric, rather than isochoric, models to relate line intensities to the radiation pressure. Model results are available for a range of values of the incident radiation field and gas pressures. The model grid ranges are $P/k_{\rm B}=n\,T_{\rm K} = 1\times10^5 - 1\times10^8$~K~cm$^{-3}$, for the gas pressure and $\chi=1$ to $1\times10^6$ for the radiation field in Mathis units.

\begin{figure}
  \centering
  \includegraphics[width=5.5cm]{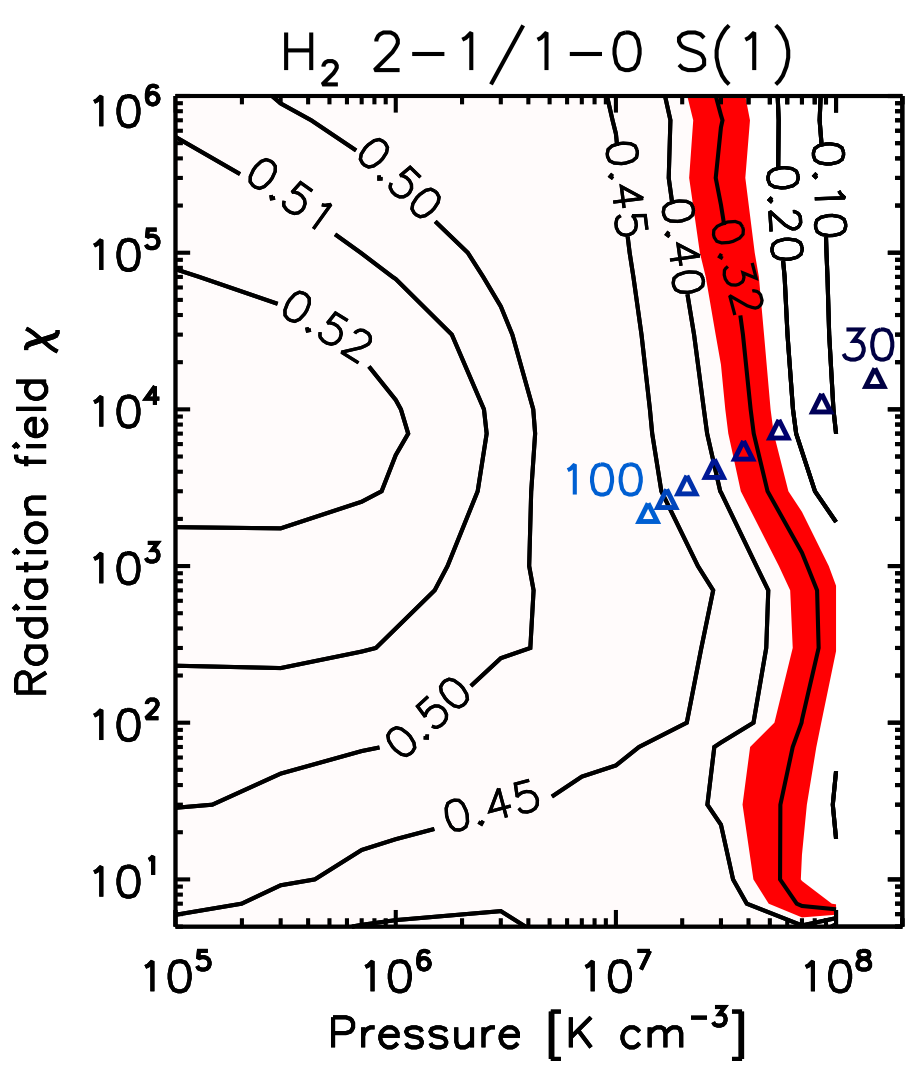}
     \caption{H$_2$~2$-$1/1$-$0~S(1) intensity line ratio estimated from the PDR Meudon code, for a grid of radiation fields and gas pressures. The measured value for the low velocity component is 0.36 which is  highlighted in the figure. The red area marks 1-$\sigma$ uncertainty associated with this value. Triangles represent the values given in Table~\ref{tab:hii}, estimated from the \hii/PDR model for radii from 30 to 100~pc.}
        \label{fig:CO_PDR}
  \end{figure}

Fig.~\ref{fig:CO_PDR} shows the model results for the H$_2$~$2-1/1-0$~S(1) line ratio. The red filled area indicates the 1$\sigma$ uncertainty on the observed ratio for the low velocity component (Table~\ref{tab:vels}). The range of possible combinations of radiation fields and pressures given in Table~\ref{fig:CO_PDR} are marked by triangles. This figure indicates that PDR models can account for the observed intensity ratios within the constrains on physical parameters set by the spatial extent of the emission lines. In the paper, we use a value $R_{\rm in}=35$~pc which corresponds to the value observed in \brg. From the model in Fig.~\ref{fig:CO_PDR}, the best R$_{\rm in}$ value is  $\sim$~55~pc. 

\end{appendix}

\end{document}